\documentclass[preprint,flushrt]{aastex}
\usepackage{amsmath}
\usepackage{graphicx}
\usepackage{subfigure}
\usepackage{ulem}

\shorttitle{UpSco}
\shortauthors{Donaldson et al.}

\begin{document}

\title{New Parallaxes for the Upper Scorpius OB Association}

\author{J.\ K.\ Donaldson\altaffilmark{1}, A.\ J.\ Weinberger\altaffilmark{1}, 
J.\ Gagn{\'e}\altaffilmark{1, 2}, 
A.\ P.\ Boss\altaffilmark{1}, S.\ A.\ Keiser\altaffilmark{1}}
\affil{Department of Terrestrial Magnetism, Carnegie Institution of Washington, 5241 Broad Branch Rd NW, Washington, DC 20015, USA; \texttt{jdonaldson@carnegiescience.edu}}
\altaffiltext{2}{NASA Sagan Fellow}

\begin{abstract}
Upper Scorpius is a subgroup of the nearest OB association, Scorpius--Centaurus. Its young age makes it an important association to study star and planet formation. We present parallaxes to 52 low mass stars in Upper Scorpius, 28 of which have full kinematics. We measure ages of the individual stars by combining our measured parallaxes with pre-main sequence evolutionary tracks. We find there is a significant difference in the ages of stars with and without circumstellar disks.  The stars without disks have a mean age of $4.9\pm0.8$\,Myr and those with disks have an older mean age of $8.2\pm0.9$\,Myr. This somewhat counterintuitive result suggests that evolutionary effects in young stars can dominate their apparent ages. We also attempt to use the 28 stars with full kinematics (i.e.\ proper motion, radial velocity, and parallax) to trace the stars back in time to their original birthplace to obtain a trackback age. We find, as expected given large measurement uncertainties on available radial velocity measurements, that measurement uncertainties alone cause the group to diverge after a few Myr.
\end{abstract}

\clearpage

\section{Introduction}
Young associations of stars ($\sim5$--10\,Myr) are ideal laboratories for the study of star and planet formation. Many young stars harbor circumstellar disks where planet formation may still be occurring. It is important to observe the timescale in which disks dissipate in order to fully understand the process of planet formation and the time during which raw materials are still available \citep[e.g.][]{Meyer07}.  The ages of associations are often derived from the average of individual pre-main sequence track ages \citep[e.g.][]{Cohen79}.  This makes luminosity, and therefore distance, and important component to the study of disk dissipation timescales.

The Upper Scorpius OB Association (Upper Sco) is a subgroup of the Scorpius-Centaurus OB Association (Sco-Cen) with an average distance of 145\,pc \citep{deZeeuw99}.  It has an age of 5--11\,Myr \citep{Preibisch02,Pecaut12}, making it an ideal place to study star and planet formation.  Sco-Cen is likely the closest analog to the environment where the Sun formed \citep{Preibisch08}.  Upper Sco is the youngest of the subgroups;  both Upper Centaurus Lupus and Lower Centaurus Crux are older.  All together these regions cover $\sim 80^\circ$ of Galactic longitude and $\sim 40^\circ$ of latitude.Upper Sco is the smallest region, covering a projected diameter of $\sim15^\circ$ or in total $\sim$300 square degrees \citep{deZeeuw99}. Upper Scorpius and the neighboring subgroups of Sco-Cen contain thousands of low-mass stars \citep{Preibisch08} whose population has not been fully characterized.  Many surveys have identified hundreds of low mass stars in Upper Sco and Sco-Cen \citep{Walter94, Preibisch98, Preibisch99, Preibisch01, Preibisch02, Song12, Rizzuto15, Pecaut16}. \cite{Carpenter06} showed that the disks around higher-mass stars evolve faster, and that therefore, by the age of Upper Sco, more lower-mass stars have disks than higher-mass stars. Indeed, many Upper Sco stars have circumstellar disks \citep{Carpenter09,Luhman12} that are ideal for study with ALMA \citep[e.g.][]{Carpenter14,Barenfeld16}.

Distance is necessary for deriving ages of stars and crucial to confirming membership status. Yet no distances have been measured to many Upper Sco low-mass members (spectral type K and later). With Hipparcos, \cite{deZeeuw99} identified 120 members, but only four K and two M stars.  The luminosities of the low mass members are calculated using an average distance to the association when placing in a Hertzsprung-Russell (H-R) diagram. \citep[e.g.][]{Aarnio08}. When compared with pre-main sequence tracks, such H-R diagrams appear to imply a wide range of ages for the low-mass members.  In the case of Upper Sco, Monte Carlo simulations indicated that this age spread was not necessarily real taking into account the observational uncertainties. \citep{Slesnick08}. 

Kinematics, such as proper motions, are also important to understand the environment stars are formed in. The 3D spatial and kinematic structure of the association affects interactions between stars and their disks. This is important for studying external photoevaporation and stellar encounters, both of which have been hypothesized for the Solar System \citep[e.g.][]{Balog07, Olczak10}.

In this paper, we present parallaxes and proper motions of 52 potential low mass Upper Sco members. In Sections~2 and 3, we discuss the observations and data analysis.We reject four candidate members from Upper Sco and discard them from further analysis.  In Section~\ref{sub:allages}, we use the distances derived to calculate ages for the stars and compare stars with circumstellar disks to stars without disks. In Section~\ref{sub:traceback}, we attempt a traceback analysis of 28 stars using Radial Velocity (RV) measurements from the literature.

\section{Observations}\label{sub:obs}
We selected K- and M-type candidate members of Upper Sco from publicly available Spitzer searches for circumstellar disks. This yielded 22 stars with infrared excesses in \cite{Carpenter06}, plus an additional nine from \cite{Carpenter09}. We added to the sample an equal number of stars that did not show infrared excess and were of the same spectral type distribution as the disk-bearing stars.  To avoid biasing the sample by using only samples from one study, the M-type sample was augmented by additional putative members from \cite{Preibisch02} and \cite{Slesnick08}. Our final sample contained 92 stars. Limited observing time constraints ultimately allowed us to measure parallaxes for 55 stars. Of these, 31 have Spitzer-detected excesses, 21 do not, and 3 were not investigated for infrared excess.

We observed the 55 potential members of Upper Sco with the CAPSCam instrument \citep{Boss09} at the 2.5\,m du Pont Telescope at Las Campanas Observatory.  A full description of the instrument and the observing methods are given in \cite{Boss09} and \cite{Weinberger13}, but we summarize a few points here.  

Our observations were performed using the guide window (GW) of CAPSCam, which is an independently readable subarray.  The bright target star was placed in the GW, which was in the center of the full field (FF).  The GW was read out more often than the FF so that the bright target star is not saturated while we integrated for longer on the FF to get better signal-to-noise on the background reference stars.  Table~\ref{tab:obs} lists the GW and FF integration times for each target along with the dates of observation. Each target has a minimum of five epochs and a minimum time baseline of 2 years. At each epoch, observations at these integration times were repeated four times at four different dither positions for a total of 16 observations.  To improve efficiency, we obtained the observations in the no guide window shutter mode, which keeps the shutter open during GW readout.

\section{Data Analysis}\label{sub:data}
\subsection{Astrometric solution}
We used all observations to derive the position, proper motion, and parallax solution for 52 of the Upper Sco targets. The three remaining stars are possible binaries and are discussed in Section~\ref{sub:binary}. We calculated the astrometric solution iteratively using the ATPa software, which is described in detail in \cite{Boss09} and \cite{Anglada12}.  A template is created from one night to determine the positions of the reference stars. At each epoch, the reference stars in each image are compared to the template to find their relative positions.  The individual positions are averaged together for each epoch.  The apparent motion of the stars is represented by an astrometric model to obtain positions, proper motions, and parallaxes for all reference stars and the target star.  The reference frame is then reconstructed by including only stars that are stable, with an epoch-to-epoch residual rms less than 1\,mas. This process is repeated two more times. The number of reference stars are listed in Table~\ref{tab:para} and on average there are 40 stars per field.

\begin{figure}
\begin{center}
	\epsscale{1}
	\vspace{-5mm}
	\subfigure{\plotone{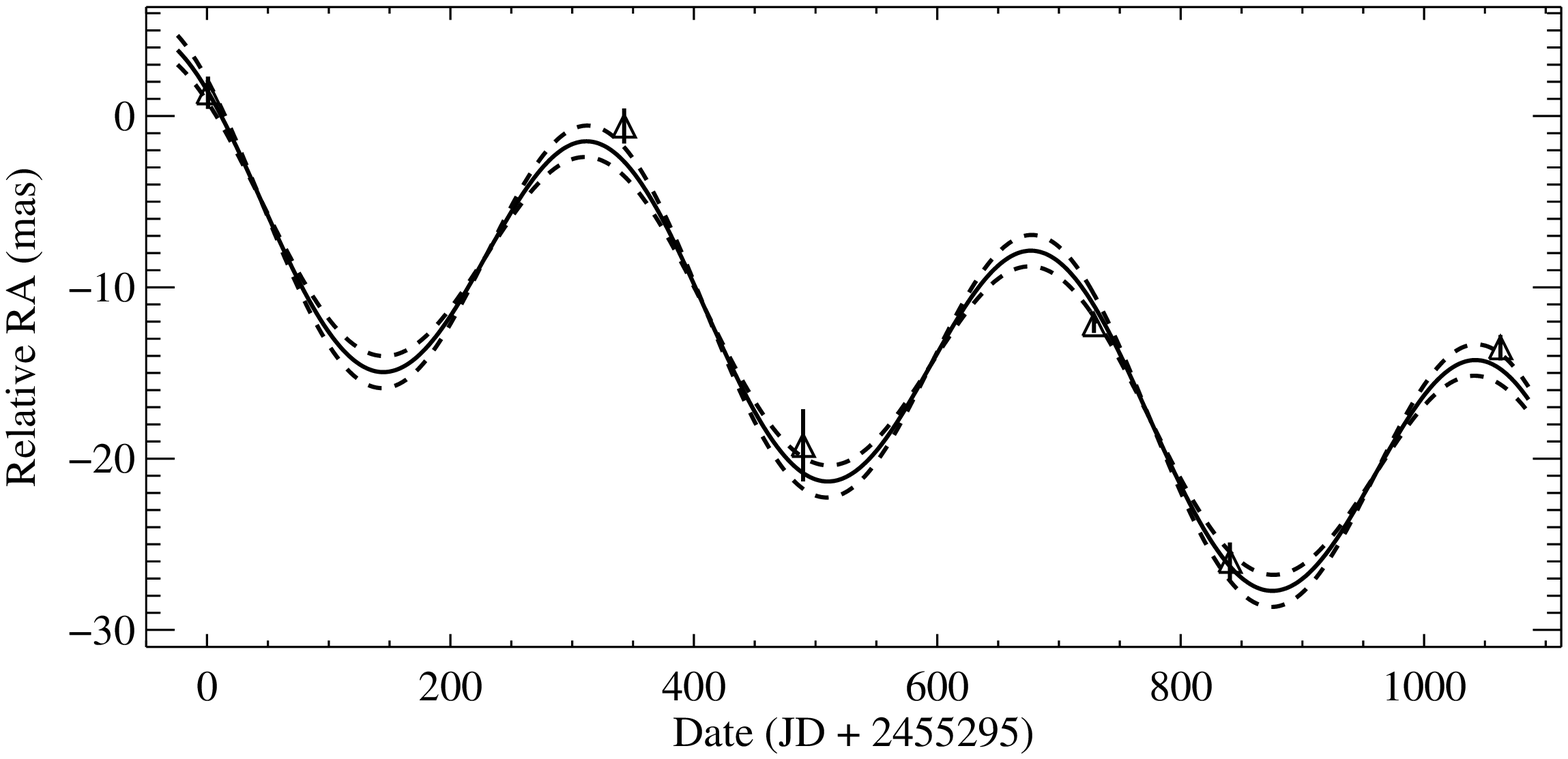}}\vspace{-5mm}
	\subfigure{\plotone{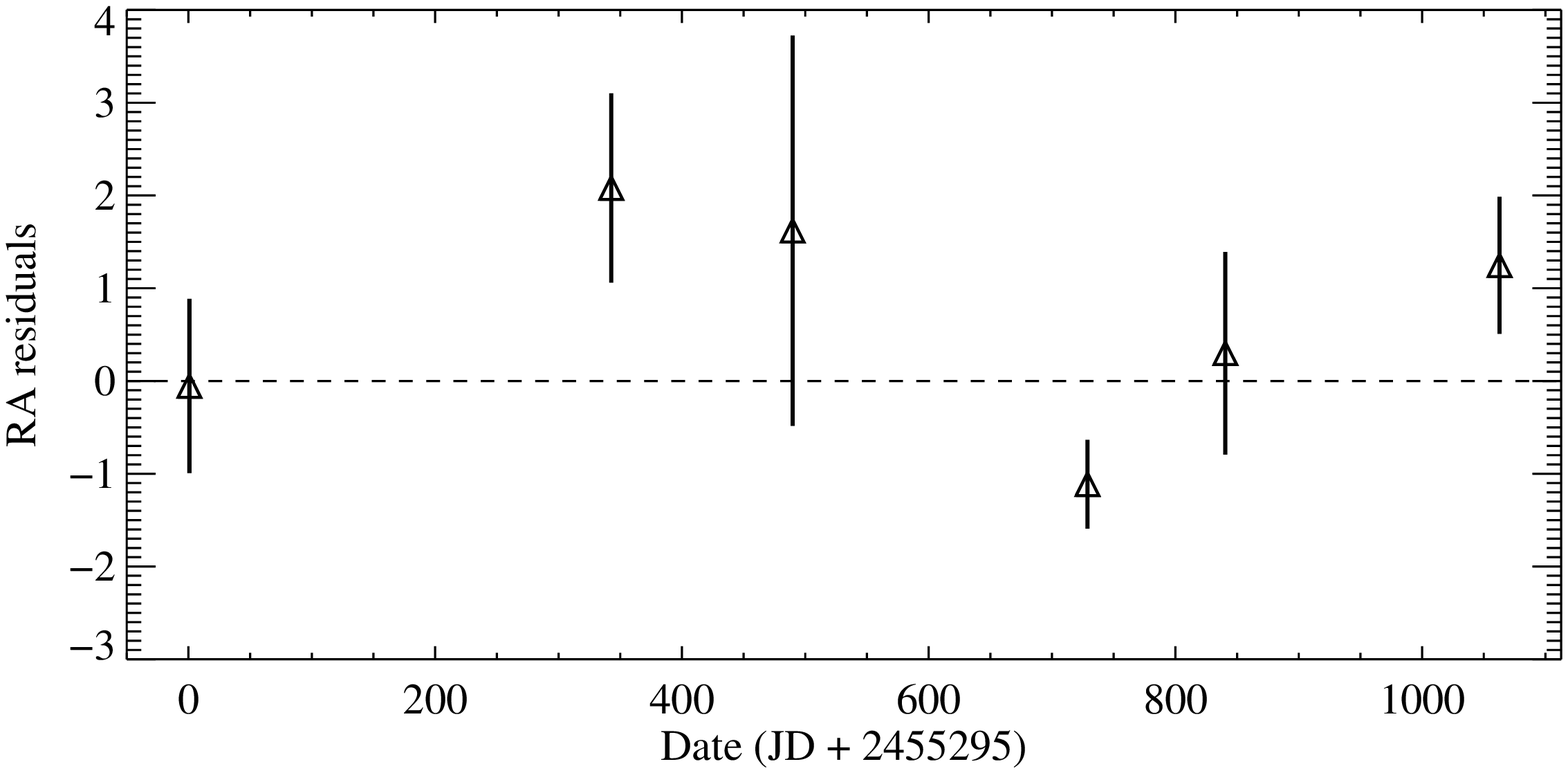}}
	\caption{The top panel shows the CAPSCam measurements of right ascension versus time of J155729.9-225843; sinusoidal parallax motion and linear proper motion are both apparent. The solid line is the best fit model for position, parallax, and proper motion, and the dashed lines show the fit at $\pm 1\sigma$ in the parallax based on our Monte Carlo trials. The bottom panel shows the residuals to the best solution. This example target is typical in that its derived parallax has an uncertainty of $\sim1\,$mas.}
	\label{fig:astro}
\end{center}
\end{figure}

\begin{figure}
\begin{center}
	\plotone{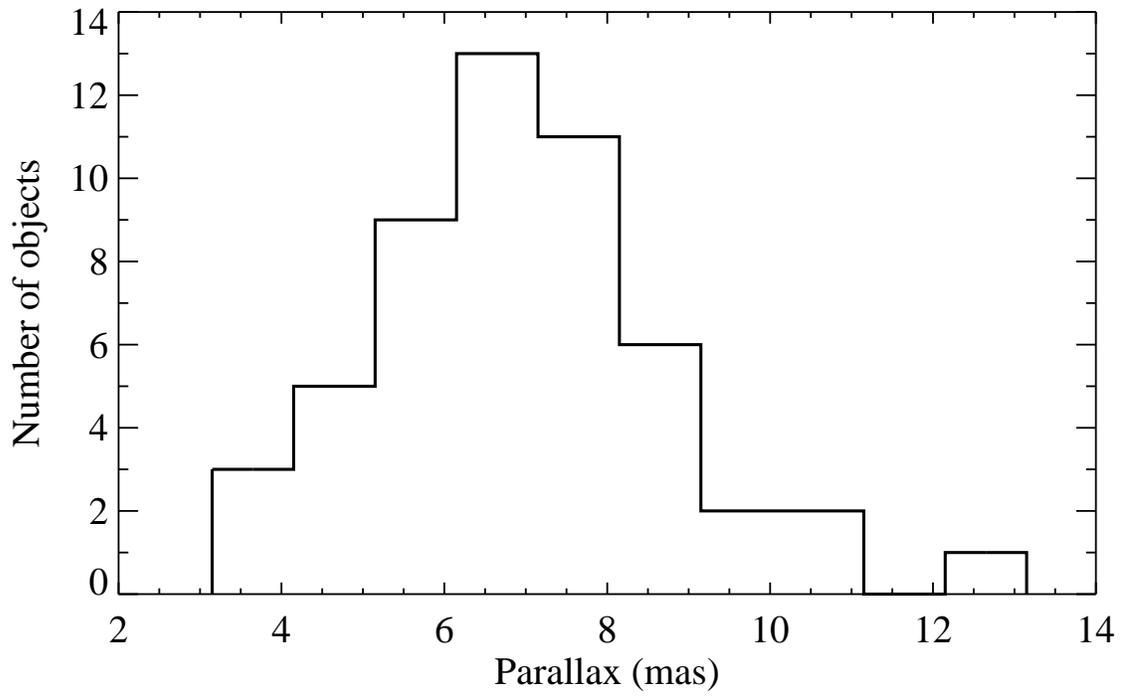}
	\caption{Histogram of the parallaxes of all 52 target stars with parallax results.}
	\label{fig:histpara}
\end{center}
\end{figure}

\begin{figure}
\begin{center}
	\plotone{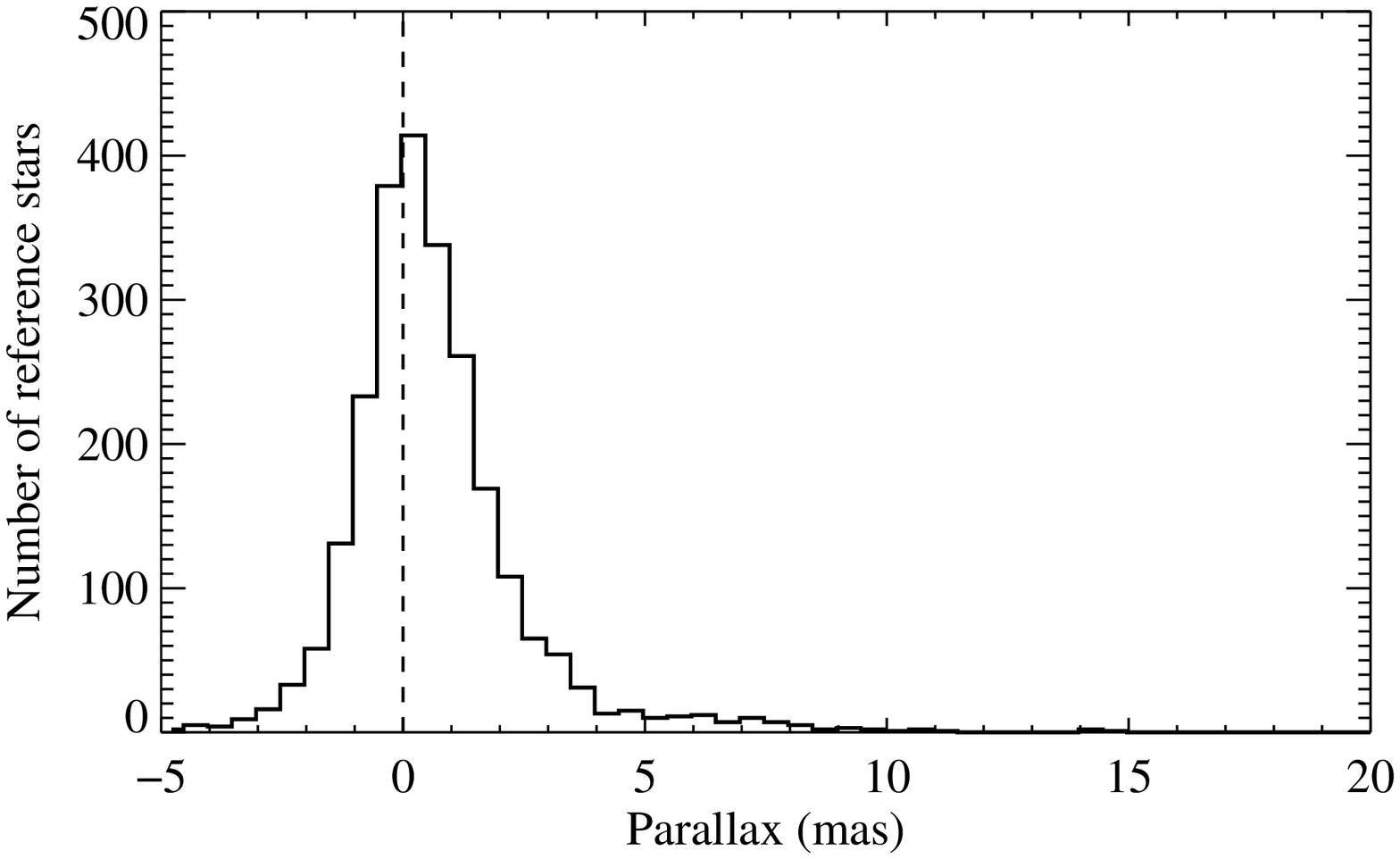}
	\caption{Histogram of the parallaxes of all reference stars used. The dashed line is at zero. The negative side of the distribution has a standard deviation of 0.88 mas. The uncertainty in the parallax distribution of the distant reference stars is therefore comparable to our Monte Carlo inferred uncertainties on the parallax for each of our targets.}
	\label{fig:refstars}
\end{center}
\end{figure}

We use a Monte Carlo method to estimate the parallax and proper motion uncertainties. We fit the starting stellar position, parallax, and proper motion in each trial. In every trial, the stellar position at each epoch is randomly drawn from a normal distribution with a mean and uncertainty measured from the individual images during the astrometric iterative solution described above. For parallax fits with chi-square greater than one, we increase the positional uncertainties at each epoch and refit until chi-square equals one.  This added uncertainty, or jitter, comes from systematic uncertainties in the data.  The uncertainty on the parallax and proper motions in Table~\ref{tab:para} comes from the standard deviation of the fitted values across all trials after adding the jitter. Table~\ref{tab:params} lists the $\Delta$\,parallax~factor, or the difference between the maximum and minimum parallax factors covered by the data, as well as the positional jitter added to each star. Figure~\ref{fig:astro} shows a typical astrometric solution as well as the residuals of the final fit. Figure~\ref{fig:histpara} shows a histogram of the parallaxes of our target stars. The distribution has a standard deviation of 1.8\,mas. Figure~\ref{fig:refstars} shows the histogram of the parallaxes of all the reference stars used. The negative side of the distribution is indicative of the uncertainties we expect from our fitted parallaxes.  The negative side of the distribution has a 1\,$\sigma$ width of 0.88 mas. The mean uncertainty in our target stars from Table~\ref{tab:para} is 1.1 mas. The uncertainties as determined by our Monte Carlo for each target thus seem reasonable. Comparisons between CAPSCam results and other programs have been shown in \cite{Weinberger16} and \cite{Donaldson16}.

\subsection{Zero-point Correction}
Since the astrometric solution is derived in comparison to the other stars in the field, the motion of reference stars can introduce a bias into the parallax solutions. To correct this effect, we fit the photometric distances to the reference stars using $B$ and $I$ band photometry from the USNO-B1 catalog and $J$, $H$, and $K_s$ photometry from the 2MASS catalog. We fit the photometry with Kurucz stellar atmosphere models.  Giant stars are recognized as such by their small distances when fit as dwarf stars, and are refit as giant stars.  Dwarf stars are only used as references if their effective temperatures are $>3800$\,K because the Kurucz models are more reliable at these temperatures.  

We then calculate the bias by measuring the average difference between the photometric parallax and the astrometric parallax for each reference star with good photometry.  We subtract the bias from the astrometric solution to get the absolute parallax.  Table~\ref{tab:para} lists the relative parallaxes, zero-point corrections, and absolute parallaxes for all stars with an astrometric solution.  More detail on this method can be found in \cite{Anglada12} and \cite{Weinberger13}.

\subsection{Proper motion offset}\label{sub:pm}

\begin{figure}
\begin{center}
	\epsscale{0.7}
	\vspace{-5mm}
	\subfigure{\plotone{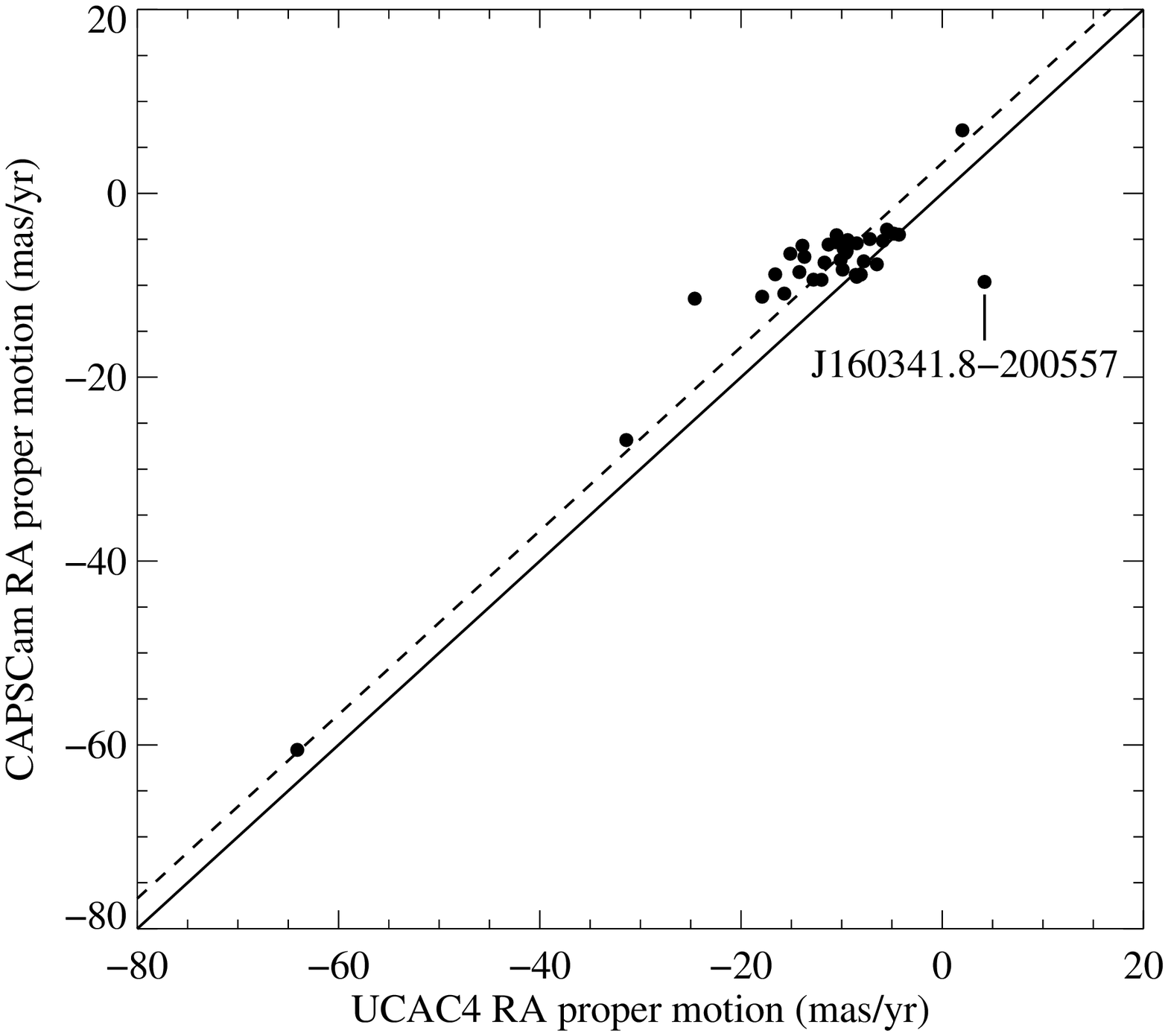}}\vspace{-5mm}
	\subfigure{\plotone{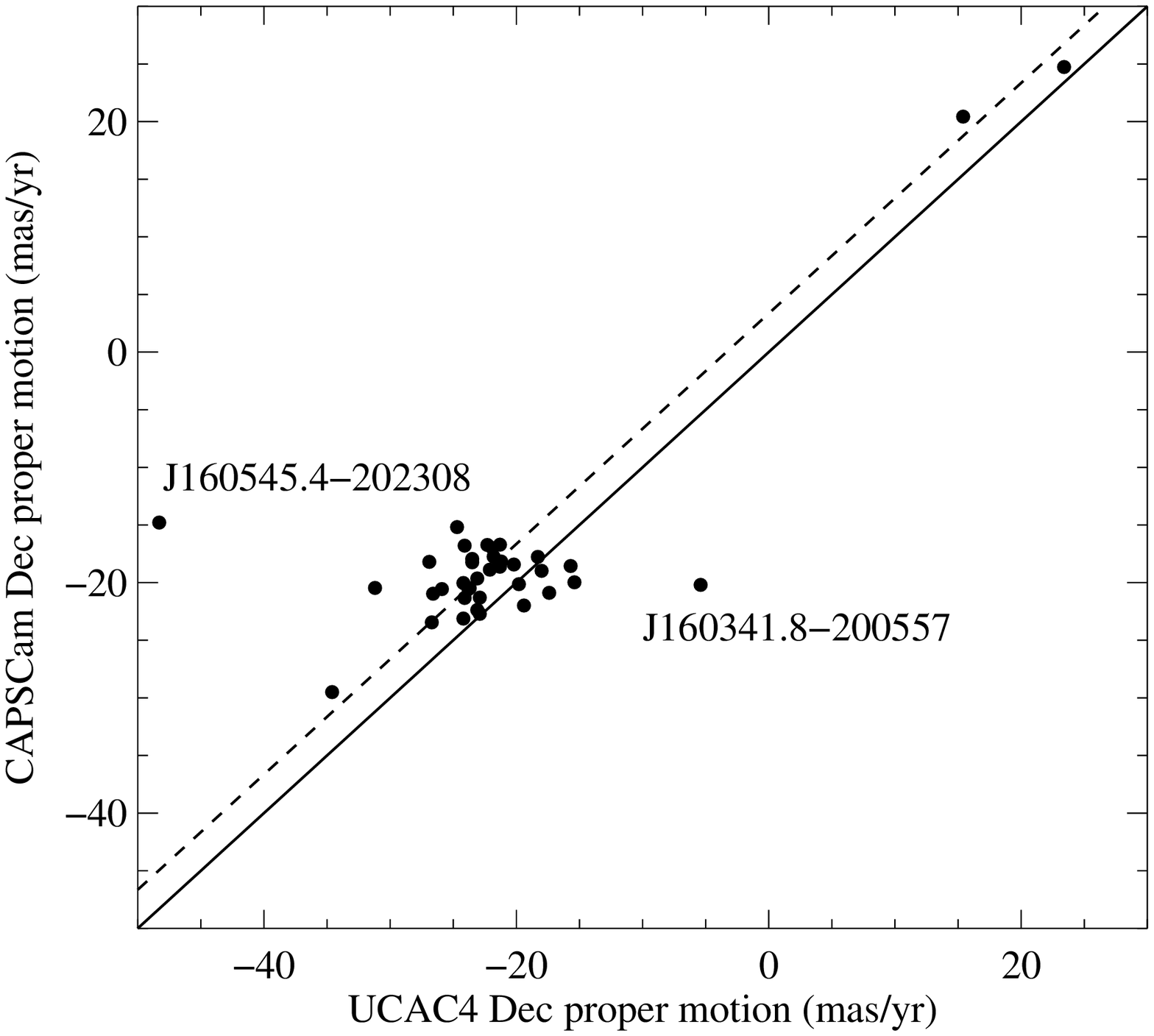}}
	\caption{Comparison of the proper motions of 38 stars in our sample with UCAC4 measurements \citep{Zacharias13}. The solid line shows a one-to-one correlation, and the dashed line represents the fit for the offset between the two datasets. The mean offset is  $3.3\pm2.8$\,mas\,yr$^{-1}$ in R.A. and $3.4\pm3.0$\,mas\,yr$^{-1}$ in declination. Errorbars are not shown for clarity, but a typical UCAC4 uncertainty in our sample is $2.8$\,mas\,yr$^{-1}$. The largest outliers are J160341.8--200557 and J160545.4--202308. See Section~\ref{sub:pm} for more details.}
	\label{fig:pm}
\end{center}
\end{figure}

Our proper motions need to be calibrated for the same reason mentioned above. Unfortunately, our reference stars generally do not have cataloged proper motions.  We compared our proper motions for the Upper Sco stars with those measured in UCAC4 \citep{Zacharias13}.  Figure~\ref{fig:pm} shows the comparison in right ascension (R.A.) and declination (decl.) of 38 of our targets that have UCAC4 measurements.  There is an offset between our measurements and those of UCAC4 of $3.3\pm2.8$\,mas\,yr$^{-1}$ in R.A. and $3.4\pm3.0$\,mas\,yr$^{-1}$ in decl.  We used these offsets to correct the proper motions of the remaining stars without UCAC4 data and propagate the uncertainties. Table~\ref{tab:para} reports the uncorrected proper motions we measured.

\subsection{Possible Binaries with no Astrometric Solutions}\label{sub:binary}

\begin{figure}[t!]
\begin{center}
	\includegraphics[width=\textwidth]{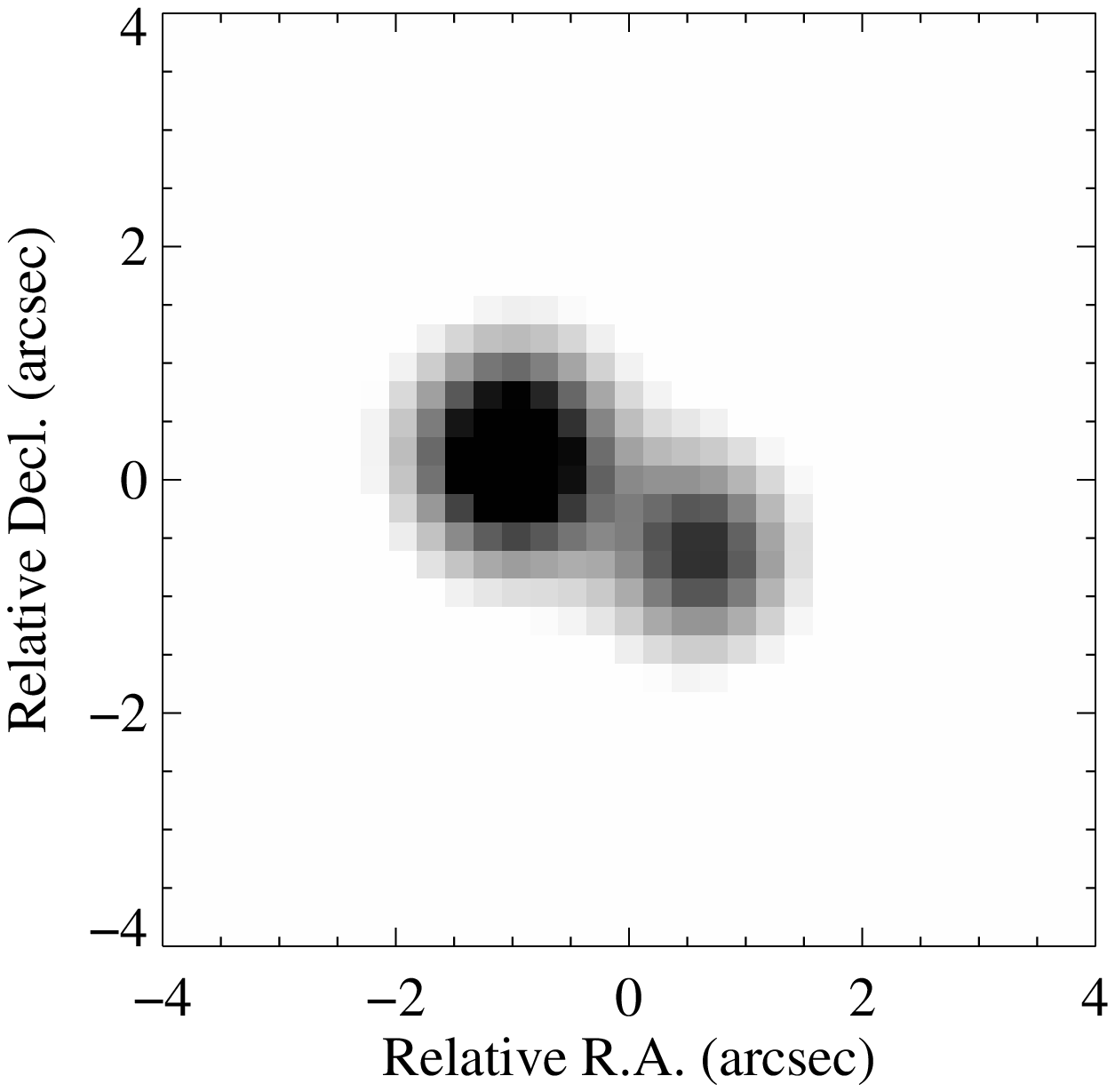}
	\caption{Image of J160702.1-201938 as observed on August 3, 2015. The binary has a visual separation of 1.39 arcsec, a position angle of 117$^{\circ}$, and a flux ratio of 0.60.  See Section~\ref{sub:binary} for more details.}
	\label{fig:binary}
\end{center}
\end{figure}

For three out of our 55 targets, ATPa was not able to converge on an astrometric solution.  One of these sources, USco J160140.8-225810, was reported as a binary by \cite{Buoy06}, and the source is elongated in our images.  One other target, J160702.1-201938, was previously reported to be a binary by \cite{Kraus07}, and is also clearly elongated or resolved in our images, as shown in Figure~\ref{fig:binary}.  In the August 3, 2015 observations, the binary has a visual separation of $1.39\pm0.02$\,arcsec, a position angle of 117$^{\circ}$, and a flux ratio of 0.60. 
The remaining source, J160719.7-202055, is not visual binary within the resolution of CAPSCam, but ATPa was still unable to converge on a solution. The star may be a binary. 

\subsection{Determining membership probabilities}\label{sub:banyan}
In order to assess membership of our targets, we used BANYAN~$\Sigma$ \citep[J.~Gagn{\'e} et.\ al.\ in preparation; see also][]{Gagne17}, which is an updated version of the BANYAN~II tool \citep{Gagne14,Malo13}.    
BANYAN-$\Sigma$ uses BayesÕ theorem with the position and kinematics of a star to determine the probability that it belongs to the Galactic field population or 21 associations younger than 1\,Gyr and within 150\,pc of the Sun. Each association is modelled with a multivariate gaussian probability distribution in $XYZUVW$ space, whereas the field is modelled with a mixture of multivariate gaussians, following the spatial and kinematic distributions of the Besan\c{c}on Galactic model \citep{Robin2012}. Sky position and proper motion are required for the analysis, but radial velocity and parallax are facultative and are marginalized analytically when no measurements are available. BANYAN~$\Sigma$ recovers a larger number of true association members compared to BANYAN~II at a fixed rate of false positives, and it includes Upper Scorpius, which was not modelled in BANYAN~II.

Using BANYAN~$\Sigma$, all but five stars in our sample obtain an Upper Sco membership probability larger than 99\%. One of the five others, J160159.7--195219, has a greater than 95\% probability, and we thus also consider it as a member. The remaining four stars, Usco--155744.9--222351, USco--160004.3--223014, Usco--160325.6--194438, and J16211564-24361173, all have membership probabilities below 2\%, and we therefore removed them from our sample in all further analysis.

\section{Pre-main sequence isochronal ages}\label{sub:allages}
\subsection{Calculating the ages of individual stars}\label{sub:ages}

To determine the ages of individual stars in our sample, we use the models of \cite{Baraffe15} to compare our absolute magnitudes with theoretical isochrones. First, we calculate absolute magnitudes in $J$, $H$, and $K$ from 2MASS using our parallax measurements. We account for extinction using A$_V$ coefficients from the literature, and convert them to 2MASS bands as following: A$_J=0.282$~A$_V$, A$_H = 0.175$~A$_V$, and A$_K = 0.112$~A$_V$ \citep{Cambresy02}.  All but two of our sources have an A$_V$ measurement from the literature, and their values are listed in Table~\ref{tab:ages} along with our computed $H$ band absolute magnitudes ($H_{\rm abs}$).

To estimate effective temperatures, we used the spectral type-temperature relations in Table~6 of \cite{Pecaut13}.  Uncertainties in spectral types are propagated to temperature measurements by converting the upper and lower bounds of the spectral type uncertainties into temperatures and computing the range.  All spectral types, effective temperatures and uncertainties are listed in Table~\ref{tab:ages}. 

Individual ages were calculated by interpolating the models of \cite{Baraffe15}. The uncertainty in the fit was derived from uncertainties in both effective temperature and parallax.  Some of the age uncertainties are relatively large ($\gtrsim 50$\%), which is a consequence of uncertainties in spectral types. This is why a constant uncertainty in effective temperature should not be assumed.  

\begin{figure}[t!]
\begin{center}
	\includegraphics[width=0.9\textwidth]{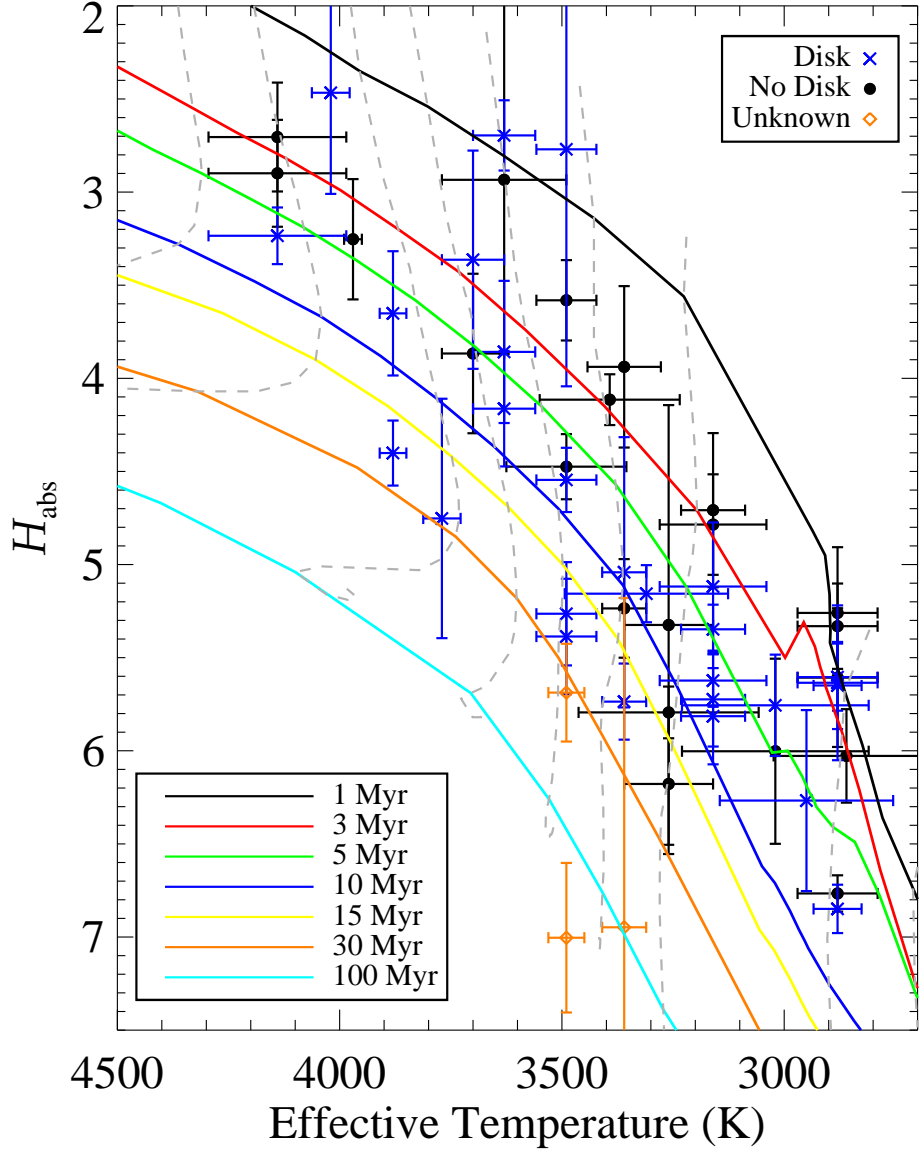}
	\caption{Absolute $H$ magnitude versus effective temperature for the Upper Sco stars in our sample.  The stars are separated into those with cirumstellar disks (blue x symbols) and those without disks (black circles). The orange diamonds represent stars where the presence of a disk is not known. The models of \cite{Baraffe15} are displayed as colored lines. The grey dashed lines represent the constant masses. From left to right they are 1.0, 0.8, 0.6, 0.5, 0.4, 0.3 0.2, and 0.06 M$_\odot$. See Section~\ref{sub:ages} for more details.}
	\label{fig:Hmag}
\end{center}
\end{figure}

Figure~\ref{fig:Hmag} shows the absolute magnitude in $H$ in our sample versus effective temperature. For comparison, 1--100\,Myr isochrones from \cite{Baraffe15} are also shown.  The gray dashed lines show constant stellar masses. The data is split into two groups, those that have a circumstellar disk, characterized by an excess in the infrared above the photosphere, and those that do not have a disk.  

After removing the four stars with low probability of membership, we calculated the mean age of our Upper Sco stars by bootstrapping, a resampling technique we use to get a better estimate of the variance. We get a mean age of $6.6\pm0.6$\,Myr and a median age of 5.2\,Myr.

Several of the stars kinematically rejected for membership in Section~\ref{sub:banyan} are significantly older than the mean age we derive.  USco--160004.3--223014 and Usco--160325.6--194438 have ages of $89\pm44$\,Myr and $79\pm43$\,Myr respectively.  This is another indication that these stars are not members of Upper Sco.

\subsection{Ages of the disk-bearing and disk-free members}\label{sub:disk}

\begin{figure}[t!]
\begin{center}
	\includegraphics[width=\textwidth]{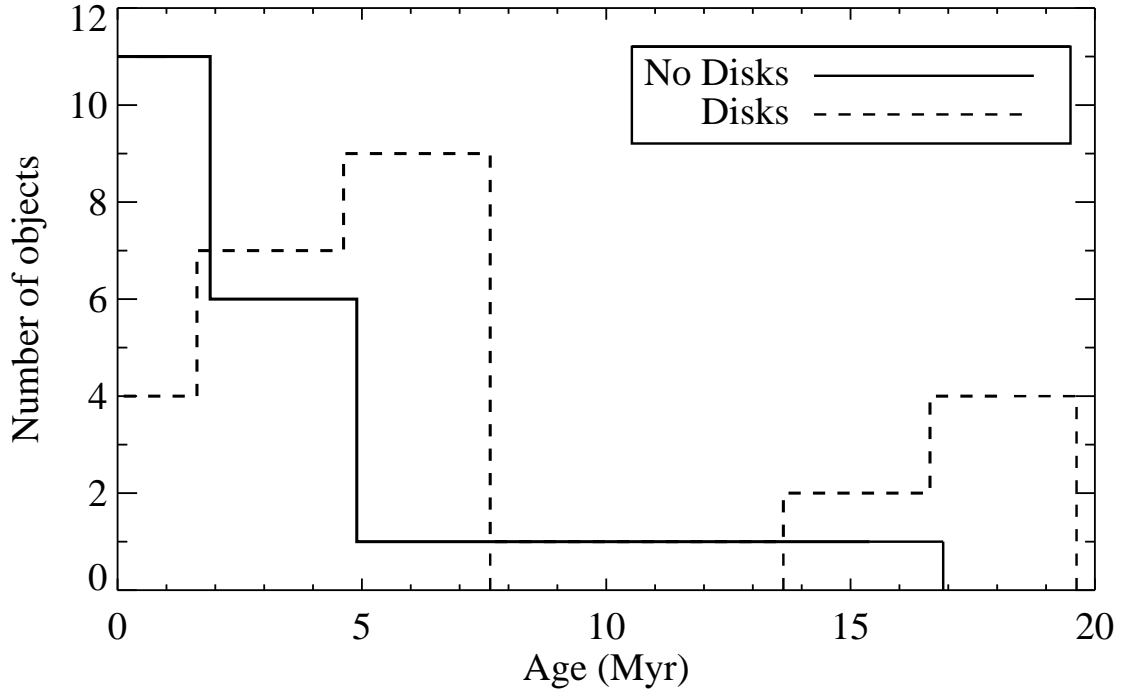}
	\caption{Histogram of the ages of the individual stars, split into those with circumstellar disks (dashed line) and those without disks (solid line). The groups appear to be two different populations. See Section~\ref{sub:disk} for more details.}
	\label{fig:histoages}
\end{center}
\end{figure}

Since many of the Upper Sco stars have disks, we should determine if there is a difference between stars with disks and those without.  Since disks disappear with time, the expectation is that stars with disks should be the same age or younger than those without disks. Figure~\ref{fig:histoages} shows a histogram of the ages of our sources broken down into stars with and without disks.  The populations are not normally distributed, and the stars with disks appear to be older on average than those without disks, in opposition to the expectation.

We separated the two groups and calculated the average of each group with the bootstrapping method.  The diskless stars have a mean age of $4.9\pm0.8$\,Myr and a median of 3.4\,Myr. The stars that have disks have a larger mean age of $8.2\pm0.9$\,Myr and a median age of 6.5\,Myr.

To confirm that the difference in ages between the two groups is significant, we used three statistical tests to determine the probability that the two groups came from the same distribution.  All three tests showed low probabilities.  The tests we used were, the Kolmogorov-Smirnov test, which yielded a probability of 0.009 of the samples coming from the same inherent population, the Student-t test, with a P-value of 0.04, and the Wilcoxon rank sum test, with a probability of 0.008.  All tests have probabilities below 5\%, so we reject the hypothesis that the ages of the disk-bearing and diskless stars are drawn from the same distribution. 

We also tested whether the two groups follow different spectral type distributions.  The two groups have almost the same distribution of spectral types, since the sample was created to ensure the two groups could be properly compared to one another.  The stars with disks have a mean spectral type of M2 with a range of spectral types from K2 to M5. The stars without disks have a mean spectral type of M3 and a range from K5 to M5. A Kolmogorov-Smirnov test gives a 73\% probability that the distributions were drawn from the same inherent population.

\begin{figure}[t!]
\begin{center}
	\includegraphics[width=\textwidth]{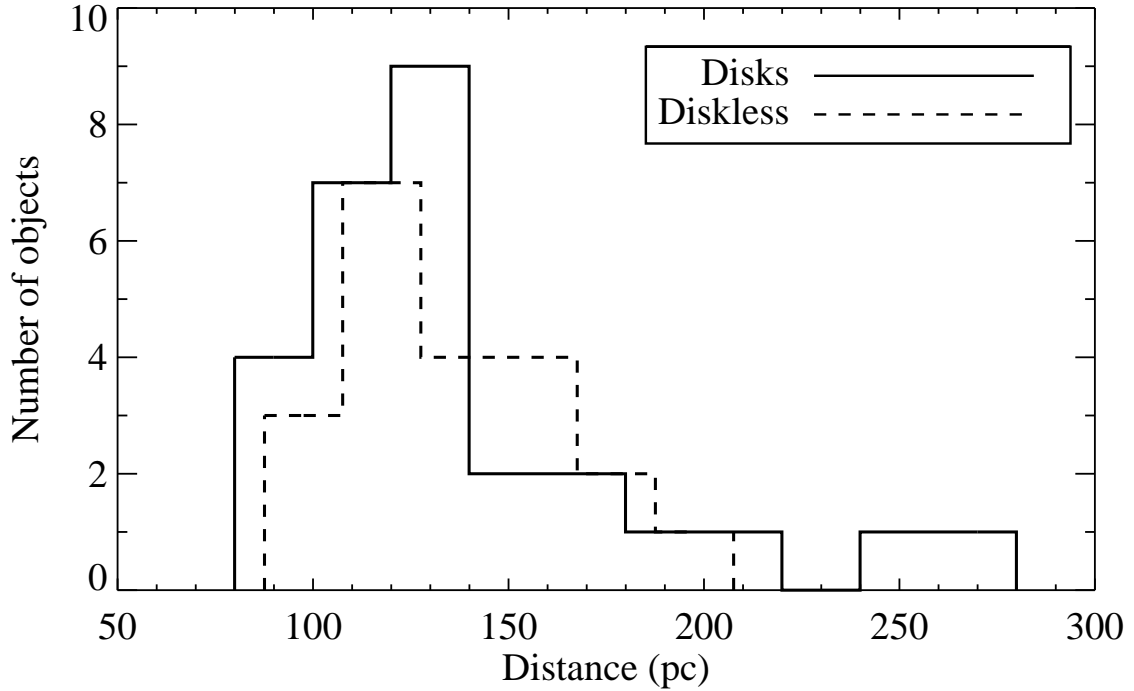}
	\caption{Histogram of the distances for the stars with disks (solid line) and those without disks (dashed line). The distributions are similar. See Section~\ref{sub:disk} for more details.}
	\label{fig:histodist}
\end{center}
\end{figure}

In Figure~\ref{fig:histodist}, we show the distribution of distances for the two populations. They look similar except for the tail at large distances for stars with disks. A Kolmogorov-Smirnov test gives a 66\% probability that they distributions are the same.  If we remove stars with distances greater than 200 pc, then the difference in the ages between the two groups remains the same (4.4\,Myr for diskless stars and 8.0\,Myr for disk stars).

If the disks were extinguishing the central star, it would lower their H band absolute magnitude, making them look older. In order for the disk to extinguish enough light from the central star, the disk would have to be edge-on, which is unlikely to be the case for the majority of disks. Plus, extinction in the H band is 0.175 times the extinction in the visible. We have visible extinctions for each star, listed in Table~\ref{tab:ages}, which would include any extinction from a disk. They range from 0 to 2.3 mag and have an average of 0.8 mag.  The mean extinction of the disk group is actually lower than the non-disk group (0.8 mag vs 1.1 mag). We don't believe that extinction is driving this effect.

One explanation for the difference between stars with disks and stars without disks comes from \cite{Baraffe09}, where they explain the spread in luminosity in H-R diagrams for low-mass stars.  Their models showed that an accreting object at a few Myr has a smaller radius and luminosity, which makes it look older than a non-accreting object.  Applied to our sources, if the stars with disks are accreting their disks or have had recent accretion, then they would appear older than the stars with no disks to accrete.  In that case, the diskless stars are a better sample to calculate the true age of Upper Sco.

Another explanation is that magnetic fields play a part. \cite{Feiden16} has argued that magnetic fields influence the derived ages of stars in Upper Sco. He claims lower mass stars are more influenced by magnetic fields, which inhibit convection and slows contraction. Hence, stars with greater influence from magnetic fields would have larger radii and higher luminosities, making them look younger.  In order to explain the discrepancy between stars with and without disks, one population, namely those without disks, would have to have a larger influence from magnetic fields. This might indicate that stars with higher magnetic fields tend to lose their disks faster.  

\subsection{Comparison with previous studies}
The age of the diskless stars, $\sim$5\,Myr, is consistent with previous studies of K and M stars, such as \cite{Slesnick08} and \cite{Pecaut16}. But this is half of the age derived for F-type stars \citep[11\,Myr;][]{Pecaut12}. \cite{Pecaut16} showed that there is an effective temperature trend with age, where the lower temperature stars appear much younger than the hotter stars. For this reason they conclude the low mass stars are not good indicators of age.  

\cite{Feiden16} used magnetic field models to claim a consistent age for Upper Sco of 10\,Myr.  His models show that an age of 5\,Myr is obtained without magnetic fields, but 10\,Myr with them. We refit our models with the \cite{Feiden16} magnetic models and we also obtain a mean age of $10.5\pm1.0$\,Myr. We used the magnetic models with a range of magnetic field strength from 22-27~kG (see \cite{Feiden16} for more details).
We still see the discrepancy between stars with disks and those without; those with disks have a mean age of $12.3\pm1.5$\,Myr and those without have a mean age of $8.9\pm1.4$\,Myr. We also used his standard models without magnetic fields as a check and got a mean age of $7.1\pm1.0$\,Myr for stars with disks, $5.6\pm1.0$\,Myr for those without disks, and $6.2\pm0.7$ for all stars. 

\section{Traceback analysis}\label{sub:traceback}

\begin{figure}[t!]
\begin{center}
	\includegraphics[width=\textwidth]{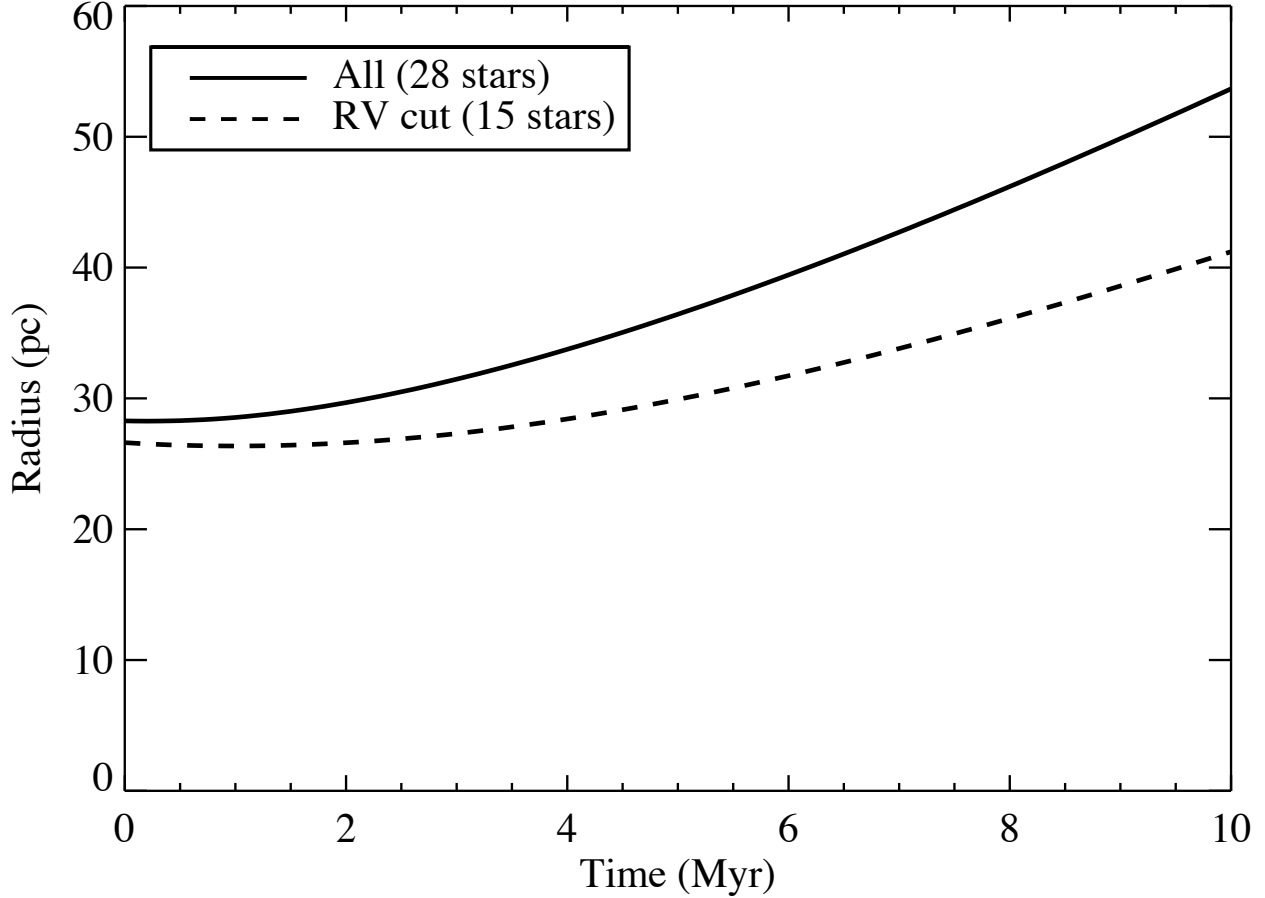}
	\caption{Radius of Upper Sco as it is traced back in time from the present to 10 Myr ago. The curves are the average radius of all the Monte Carlo trials.  Two tracebacks were run, one with all 28 stars with full kinematics (solid line) and one with only the 15 stars with better than 1\,km\,s$^{-1}$ RV precision (dashed line). Both groups diverge.  See Section~\ref{sub:traceback} for more details.}
	\label{fig:tracerad}
\end{center}
\end{figure}

The traceback method allows us to determine the age of a cluster using the present day positions and velocities to trace stars back in time to find a convergent point where they occupy the smallest volume.  The time in the traceback when the cluster has the smallest size would then be the age of the association.  This should help avoid confusion in ages between stars with and without disks since the presence of a disk should not affect the traceback. It is also independent of pre-main sequence evolution models and effective temperature trends with age.

Full 3D kinematics are needed to perform a traceback analysis. In addition to our parallax and proper motions, we included RVs from \cite{Dahm12}. This gives us RVs for 28 stars in our sample.  

Using these 28 stars, we traced them back in time, assuming their motions are straight lines, by subtracting each star's velocity multiplied by time from its present position over the course of 10\,Myr.  At each time step of 0.1\,Myr, we calculated the radius of the association as the average distance of each star from the central mean position. The time when the stars have the smallest radius is determined to be the age.

To account for the uncertainties in our measurements, we used a Monte Carlo method. We ran 10,000 trials, and in each trial we included an uncertainty term in our parallax, proper motion, and RV values. This term consists of a random number drawn from a normal distribution, scaled by the measured uncertainty for that value.  This method captures the range of possible values the true motions and distance could have. Our final output is the average radius at each time step over all 10,000 trials.

The solid line in Figure~\ref{fig:tracerad} shows the radius of the 28 stars as they go from present time (time zero) to 10\,Myr ago.  The traceback radius diverges rather than converging as expected.

\cite{Donaldson16} and \cite{Riedel17} showed that large uncertainties in position and velocity, particularly RV, can add so much noise to the Monte Carlo trials that a true convergence is impossible to find.  Some of the RV uncertainties from \cite{Dahm12} are large and may throw off the traceback analysis. To account for this, we selected only stars with better than 1\,km\,s$^{-1}$ RV precision with uncertainties ranging from 0.16-0.91\,km\,s$^{-1}$ and an average of 0.5\,km\,s$^{-1}$. This leaves 15 stars in the sample.  We ran the trackback with the new sample, and the resulting radius over time is plotted in the dashed line in Figure~\ref{fig:tracerad}. The resulting traceback still diverges. 

\cite{Donaldson16} showed in the TW Hya association that RV measurement precision needs to be very precise to get an accurate traceback result. If the velocity dispersion of the cluster is low, the stars won't spread far enough apart to allow for a good traceback. Given a typical velocity dispersion for a cluster \citep[1-2\,km\,s$^{-1}$;][]{Adams00}, \cite{Donaldson16} showed that an RV precision of  $<0.25$\,km\,s$^{-1}$ was necessary to accurately traceback the TW Hya association to within 1 Myr, assuming {\it Gaia} precision for parallax and proper motion. \cite{Riedel17} also showed with their traceback analysis that the tracebacks are limited by precision and will tend to diverge unless measurement precision is high. For Upper Sco, an RV precision of 1\,km\,s$^{-1}$ is not sufficient for an accurate traceback.

\section{Summary}
We measured the parallaxes and proper motions of 52 potential Upper Sco members. All but four have high probability of membership. Of the high probability stars, 28 have RV measurements, and therefore their full kinematics are presented.

We showed that stars with disks have an older mean isochronal age than stars without disks. This unexpected result suggests that evolutionary effects in young stars can effect their apparent ages. We also showed that current measurement uncertainties in Upper Sco are too large for an accurate traceback analysis, which might be unsurprising considering the work of \cite{Donaldson16} and \cite{Riedel17}.

\vspace{2cm}We would like to thank the staff of the Las Campanas Observatory and especially the operators of the du Pont telescope. Guillem-Anglada Escud{\'e} provided data reduction assistance and advice. This work was supported by the National Science Foundation under Grant Number 1313043. This work makes use of the Simbad database and Vizier catalogue access tool, CDS, Strasbourg, France, and the 2MASS survey, which is a joint project of the University of Massachusetts and the Infrared Processing and Analysis Center/California Institute of Technology, funded by the National Aeronautics and Space Administration and the National Science Foundation.

\begin{deluxetable}{lccl}
\tablewidth{0pt}
\rotate
\tablecaption{Record of Observations}
\tabletypesize{\scriptsize}
\tablehead{
\colhead{Designation} &\multicolumn{2}{l}{Integration Times} & \colhead{Epochs} \\
 \cline{2-3}	& \colhead{FF\tablenotemark{a}} & \colhead{GW\tablenotemark{b}} \\
			& \colhead{(s)} & \colhead{(s)} & \colhead{(JD)} \\
}
\startdata
J155106.6-240218 & 30 & 2.5 & 2455295.7, 2455367.6, 2455634.8, 2455783.5, 2456494.5, 2457178.6   \\
J155624.8-222555 & 90 & 15 & 2454992.6, 2455294.8, 2455368.6, 2455404.5, 2455634.8, 2456519.5, 2457179.6  \\
J155655.5-225839 & 45 & 2.5& 2455295.7, 2455369.6, 2455635.9, 2455783.6, 2456357.8, 2456494.5, 2456772.8    \\
J155706.4-220606 & 24 & 12 & 2455295.8, 2455406.5, 2455634.8, 2455784.6, 2456023.8, 2456135.6, 2457179.6     \\
J155729.9-225843 & 27 & 9 & 2455295.8, 2455406.5, 2455637.8, 2455784.6, 2456023.8, 2456135.6, 2456357.8   \\
Usco-155744.9-222351 & 30 & 5 & 2455296.8, 2455369.6, 2455638.8, 2455783.6, 2456520.5,  2457179.6     \\
J155829.8-231007 & 20 & 1 & 2454992.6, 2455294.8, 2455367.6, 2455406.5, 2455634.9, 2456494.5, 2457178.6, 2457238.5    \\
J155918.4-221042 & 48 & 4 & 2455295.7, 2455406.5, 2455634.9, 2455785.6, 2456023.8, 2456088.7, 2457179.6, 2457471.8     \\
USco-160004.3-223014 & 25 & 5 & 2455788.5, 2456086.6, 2456490.5, 2456772.8, 2457179.7, 1724368.2   \\
J160013.3-241810 & 40 & 4 &  2456089.6, 2456134.6, 2456490.5, 2456770.7,  2456805.7, 2457178.7, 2457238.5, 2457472.8   \\
Usco-160018.4-223011 & 25 & 5 & 2455296.8, 2455410.5, 2455788.5, 2456023.8, 2456086.7, 2456490.5, 2456772.8, 2457410.9    \\
J160108.0-211318 & 20 & 0.5 & 2455296.8, 2455369.6, 2455665.9, 2455784.5, 2456086.7, 2457179.7, 2457472.8    \\
J160159.7-195219 & 45 & 15 & 2455297.8, 2455404.5, 2455638.8, 2455785.6, 2456023.9, 2456520.5, 2457179.7   \\
J160200.3-222123 & 40 & 2 & 2456089.6, 2456135.5, 2456490.6, 2456770.7, 2456805.7, 2457178.8, 2457238.5    \\
USco-160202.9-223613 & 40 & 10 & 2455297.8, 2455406.6, 2455786.5, 2456023.9, 2456088.7, 2456357.9, 2456490.6, 2457471.8   \\
J160226.2-200241 & 40 & 10 & 2455410.5, 2456089.6, 2456135.5, 2456360.9, 2456428.7, 2456521.5, 2456771.8, 2457471.8     \\
Usco-160258.5-225649 & 45 & 3 & 2455296.9, 2455367.7, 2455637.9, 2455784.5, 2456490.6, 2457179.7   \\
Usco-160325.6-194438 & 30 & 5 & 2455297.8, 2455410.5, 2455786.5, 2456023.9, 2456088.7, 2456770.7, 2457471.8  \\
J160329.4-195503 & 40 & -- &  2455410.5, 2456089.6, 2456134.6, 2456362.9, 2456428.8, 2456521.5, 2456772.8   \\
J160341.8-200557 & 30 & 1.5 & 2455296.9, 2455369.6, 2455637.8, 2455784.5, 2456494.5, 2457179.7  \\
J160357.6-203105 & 40 & 4 & 2456089.7, 2456135.5, 2456488.5, 2456518.5, 2456770.7, 2456807.7, 2456851.6   \\
J160357.9-194210 & 30 & 6 & 2454992.6, 2455295.8, 2455368.6, 2455407.5, 2455635.9, 2456770.7, 2457238.6    \\
J160418.2-191055 & 30 & 6 & 2455297.9, 2455407.5, 2455638.8, 2455786.6, 2456024.8, 2456520.5, 2457179.7   \\
J160421.7-213028 & 24 & 1.5 & 2454992.7, 2455295.8, 2455369.6, 2455637.8, 2455788.5, 2456086.7, 2457179.7, 2457443.8  \\
J160435.6-194830 & 40 & -- &  2455410.6, 2456089.7, 2456135.5, 2456358.9, 2456488.6, 2456705.9, 2456770.8, 2456807.8, 2456851.7, 2457238.5 \\
J160439.1-194245 & 40 & 10 & 2456089.7, 2456135.5, 2456358.9, 2456488.6, 2456770.8, 2457178.7, 2457238.5   \\
J160449.9-203835 & 40 & 20 & 2455410.6, 2456089.7, 2456134.5, 2456488.6, 2456705.9, 2456770.8, 2457179.7, 2457238.5, 2457471.9   \\
J160516.1-193830 & 40 & 10 & 2455296.9, 2455407.5, 2455638.8, 2455786.6, 2456024.8, 2456088.8, 2456358.8, 2456494.6, 2457471.9   \\
J160532.1-193315 & 40 & 10 & 2455297.9, 2455407.5, 2455638.8, 2455786.6, 2456024.8, 2456088.8, 2456358.8, 2456494.6, 2457471.9   \\
J160545.4-202308 & 48 & 6 & 2455295.8, 2455407.5, 2455786.6, 2456024.8, 2456086.7, 2456520.5  \\
J160612.5-203647 & 30 & 1 & 2455295.8, 2455369.7, 2455637.8, 2455784.6, 2456494.6, 2456851.7, 2456851.7, 2457471.9  \\
J160622.8-201124 & 40 & 10 & 2455297.9, 2455369.7, 2455635.9, 2455637.8, 2455784.6, 2456494.6, 2457179.8   \\
J160643.8-190805 & 24 & 1 & 2454992.7, 2455295.9, 2455368.7, 2455407.5, 2455638.8, 2455786.6, 2456024.9, 2456357.9, 2456494.6\\ 
&&&\hspace{1.45cm}2457177.7, 2457238.6, 2457443.8, 2457472.8   \\
J160703.9-191132 & 36 & 3 & 2454992.7, 2455295.9, 2455368.7, 2455635.9, 2456024.9, 2456357.9, 2456490.6, 2457177.7, 2457238.6\\
&&&\hspace{1.45cm}2457443.8, 2457472.8   \\
J160708.7-192733 &  45 & 9 & 2456772.7,  2456850.7, 2456890.6, 2457082.9, 2457177.7, 2457237.6  \\
J160739.4-191747 & 45 & 2.5 & 2456771.9, 2456807.7, 2456890.6, 2457178.6, 2457237.5, 1723000.5    \\
J160823.2-193001 & 30 & 1 & 2455295.9, 2455407.5, 2455638.9, 2455787.5, 2456024.9, 2456518.5, 2457238.6, 2457472.8  \\
J160827.5-194904 & 40 & 10 & 2456523.5, 2456770.9, 2456807.7, 2457082.9, 2457174.7, 2457471.9   \\
J16083646-24453053 & 40 & 10 & 2456523.5, 2456771.7, 2456851.7, 2457082.9, 2457174.7, 2457237.6   \\
J160856.7-203346 & 40 & 4 & 2455369.7, 2455638.9, 2455787.6, 2456024.9, 2456488.8, 2457179.8, 2457472.9  \\
J160900.0-190836 & 40 & 5 & 2456523.5, 2456771.8, 2456807.8, 2457082.9, 2457177.6   \\
J160900.7-190852 & 40 & 2 & 2456523.5, 2456771.8, 2456807.8,  2457082.9, 2457177.7  \\
J160953.6-175446 & 60 & 15 & 2455297.9, 2455638.9, 2456024.9, 2456088.8, 2456358.9, 2456494.7, 2456521.5, 2456771.8, 2456807.7\\ 
&&&\hspace{1.45cm}2457177.7, 2457443.8, 2457472.9  \\
J160954.4-190654 & 45 & 1 & 2456555.5, 2456772.7, 2456851.7, 2457178.6, 2457236.6   \\
J160959.4-180009 & 40 & 4 & 2455297.9, 2455368.8, 2455407.6, 2455638.9, 2455787.6, 2456024.9, 2456088.8, 2456358.9, 2456490.6\\
&&&\hspace{1.45cm}2456521.5, 2456771.8, 2456807.7, 2457443.9, 2457472.9  \\
J161011.0-194603 & 40 & -- & 2455369.7, 2455638.9, 2455787.6, 2456024.9, 2456428.8   \\
J161014.7-191909 & 40 & 2 & 2456552.5, 2456772.8, 2456851.7, 2457178.7, 2457236.6, 2457471.9  \\
J161024.7-191407 & 40 & 5 &  2456553.5, 2456772.8, 2456850.6, 2457177.7, 2457235.6, 2457471.9  \\
J161052.4-193734 & 36 & 12 & 2454992.7, 2455294.9, 2455368.7, 2455404.6, 2455637.9, 2456086.7, 2456357.9, 2456494.6, 2456852.7,   \\
J161115.3-175721 & 39 & 3 & 2454992.7, 2455297.9, 2455368.7, 2455406.6, 2455637.9, 2456086.7,  2456358.9,  2456490.6,
2456771.8, 2457472.9  \\
J16211564-24361173 & 48 & 16 & 2456552.5, 2456772.9, 2456852.7, 2457178.7, 2457233.6, 2457472.9  \\
J16212490-24261446 & 60 & -- & 2456554.5, 2456772.9, 2456852.7,  2457178.7, 2457235.6, 2457472.9   \\
\enddata
\label{tab:obs}
\tablenotetext{a}{FF: Full Field}
\tablenotetext{b}{GW: Guide Window}
\tablecomments{See Section~\ref{sub:obs} for more details}
\end{deluxetable}

\begin{deluxetable}{lrrrrrrr}
\tablewidth{0pt}
\tabletypesize{\footnotesize}
\tablecaption{Parameters of astrometric fit}
\tablehead{
\colhead{Designation} & \colhead{\# of} & \colhead{$\Delta$time} & \colhead{\# of} & \colhead{$\Delta$ parallax} & \colhead{$\Delta$ parallax} 
& \colhead{Jitter} & \colhead{Jitter}\\
&  \colhead{epochs} & \colhead{(years)} & \colhead{ref stars} & \colhead{factor - R.A.} & \colhead{factor - Decl.} & \colhead{R.A. (mas)} & \colhead{Decl. (mas)}
}
\startdata
J155106.6-240218 & 6 & 5.2 & 50 & 1.86 & 0.41 & 0.0 & 11.52\\
J155624.8-222555 &  7 & 6.0 & 45 & 1.88 & 0.40 & 0.27 & 1.97\\
J155655.5-225839 &  7 & 4.1 & 38 & 1.77 & 0.36 & 0.42 & 5.25\\
J155706.4-220606 & 7 & 5.2 & 40 & 1.87 & 0.39 & 1.42 & 2.29\\
J155729.9-225843 & 7 &2.9 & 42 & 1.91 & 0.39 & 1.26 & 2.11\\
Usco-155744.9-222351 & 6 & 5.2 & 45 & 1.86 & 0.39 & 1.40 & 0.56\\
J155829.8-231007 & 8 & 6.2 &35 &  1.84 & 0.38 & 0.72 & 5.01\\
J155918.4-221042 & 8 & 6.0 & 43 & 1.87 & 0.39 & 1.83 & 2.82\\
USco-160004.3-223014 & 6 & 4.4 & 33 & 1.82 & 0.35 & 7.01 & 5.39\\
J160013.3-241810 & 8 & 3.8 & 56 & 1.41 & 0.32 & 1.61 & 1.60\\
Usco-160018.4-223011 & 8 & 5.8 & 17 & 1.67 & 0.36 & 1.60 & 1.69\\
J160108.0-211318 & 7 & 6.0 & 41 & 1.87 & 0.38 & 1.46 & 0.99\\
J160159.7-195219 & 7 & 5.2 & 39 & 1.87 & 0.37 & 0.51 & 3.97 \\
J160200.3-222123 & 7 &3.2 & 51 & 1.41 & 0.30 & 0.0 & 1.80 \\
USco-160202.9-223613 & 8 & 6.0 & 51 & 1.92 & 0.38 & 1.85 & 1.67\\
J160226.2-200241 & 8 &5.7 & 49 & 1.92 & 0.38 & 1.35 & 2.12\\
Usco-160258.5-225649 & 6 & 5.2 & 39 & 1.86 & 0.37 & 0.0 & 2.63\\
Usco-160325.6-194438 & 7 & 6.0 & 53 & 1.81 & 0.35 & 3.64 & 1.77 \\
J160329.4-195503 & 7 & 3.8 & 42 & 1.91 & 0.37 & 0.91 & 1.17\\
J160341.8-200557 & 6 & 5.2 & 38 & 1.86 & 0.36 & 0.40 & 0.99\\
J160357.6-203105 & 7 & 2.1 & 43 & 1.47 & 0.29 & 0.0 & 1.61 \\
J160357.9-194210 & 7 & 6.2 & 60 & 1.83 & 0.36 & 0.0 & 0.0\\
J160418.2-191055 & 7 & 5.2 & 36 & 1.87 & 0.36 & 1.07 & 1.53 \\
J160421.7-213028 & 8 & 5.8 & 46 & 1.94 & 0.38 & 0.0 & 1.93\\
J160435.6-194830 & 10 &5.0& 61 & 1.89 & 0.38 & 1.90 & 0.88 \\
J160439.1-194245 & 7 & 3.2 & 54 & 1.82 & 0.36 & 0.40 & 1.14\\
J160449.9-203835 & 9 & 5.7 & 51 & 1.89 & 0.37 & 2.13 & 2.51\\
J160516.1-193830 & 9 & 6.0 & 41 & 1.91 & 0.37 & 2.93 & 6.25 \\
J160532.1-193315 & 9 & 6.0 & 39 & 1.82 & 0.36 & 2.75 & 5.46\\
J160545.4-202308 & 6 & 3.3 & 68 & 1.68 & 0.32 & 1.50 & 4.51\\
J160612.5-203647 & 8 & 6.0 & 40 & 1.86 & 0.36 & 1.12 & 2.68\\
J160622.8-201124 & 7 & 5.2 & 45 & 1.87 & 0.36 & 0.66 & 3.80\\
J160643.8-190805 & 13 & 6.8 & 40&1.86 & 0.35 & 1.59 & 5.18\\
J160703.9-191132 & 11 & 5.8 &42 & 1.75 & 0.35 & 3.84 & 1.49\\
J160708.7-192733 & 6 & 1.3 & 44 & 1.96 & 0.37 & 0.0 & 0.0\\
J160739.4-191747 & 6 & 1.8 & 51 & 1.48 & 0.26 & 0.99 & 2.45\\
J160823.2-193001 & 8 & 6.0 & 49 & 1.87 & 0.35 & 1.00 & 1.64\\
J160827.5-194904 & 6 & 2.7 & 44 & 1.95 & 0.37& 1.76 & 2.80\\
J16083646-24453053 & 6 & 2.0 & 50& 1.95 & 0.37 & 1.20 & 0.0\\
J160856.7-203346 & 7 & 5.8 & 37 & 1.87 & 0.35 & 1.25 & 1.96\\
J160900.0-190836 & 5 & 1.8 & 57 & 1.95 & 0.37 & 0.62 & 2.60\\
J160900.7-190852 & 5 &1.8 & 54 & 1.95 & 0.37 & 0.97 & 3.39\\
J160953.6-175446 & 12 & 6.0 & 37&1.95 & 0.37 & 1.43 & 3.78\\
J160954.4-190654 & 5 & 1.9 & 52 & 1.39 & 0.24 & 0.0 & 0.0\\
J160959.4-180009 & 14 & 6.0 & 42 &1.92 & 0.36 & 2.85 & 2.34 \\
J161011.0-194603 & 5 & 2.9 & 49 & 1.87 & 0.35 & 0.0 & 1.37\\
J161014.7-191909 & 6 & 2.6 & 21 & 1.41 & 0.23 & 0.0 & 0.64\\
J161024.7-191407 & 6 & 2.6 & 39 & 1.75 & 0.32 & 0.0 & 4.46\\
J161052.4-193734 & 9 & 5.1 & 44 & 1.81 & 0.34 & 1.50 & 4.29\\
J161115.3-175721 & 10 & 5.8 & 43 &1.83 & 0.35 & 2.12 & 1.28 \\
J16211564-24361173 & 6 &2.5 & 45 & 1.73 & 0.29 & 1.53 & 1.24\\
J16212490-24261446 &6 &2.5 & 29 & 1.80 & 0.34 & 3.68 & 2.32 \\
\enddata
\label{tab:params}
\tablecomments{See Section~\ref{sub:data} for more details}
\end{deluxetable}

\begin{deluxetable}{lrrrrrr}
\tablewidth{0pt}
\tabletypesize{\footnotesize}
\tablecaption{Measured Parallaxes and Proper Motions}
\tablehead{
\colhead{Designation} & \colhead{$\pi_\text{rel}$}  & \colhead{Zero Point} & \colhead{$\pi_\text{abs}$} & \colhead{Distance} 
& \colhead{$\mu_\text{ra}$} & \colhead{$\mu_\text{dec}$} \\
			 & \colhead{(mas)}  & \colhead{(mas)} & \colhead{(mas)} & \colhead{(pc)} &
			  \colhead{(mas\,yr$^{-1}$)} & \colhead{(mas\,yr$^{-1}$)}
}
\startdata
J155106.6-240218 & 8.09$\pm$0.55 & -0.16$\pm$0.37 & 8.25$\pm$0.66 & 121.1$\pm$
9.75 & -10.91$\pm$0.35 & -20.14$\pm$0.47  \\
J155624.8-222555 & 7.59$\pm$0.46 & -0.34$\pm$0.15 & 7.93$\pm$0.48 & 126.1$\pm$
7.68 & -8.183$\pm$0.13 & -20.81$\pm$0.22 \\
J155655.5-225839 & 7.70$\pm$0.68 & 0.72$\pm$1.2 & 6.98$\pm$1.4 & 143.2$\pm$28.2
 & -9.390$\pm$0.17 & -21.33$\pm$0.19 \\
J155706.4-220606 & 6.63$\pm$0.84 & -0.81$\pm$0.22 & 7.44$\pm$0.87 & 134.4$\pm$
15.7 & -6.845$\pm$0.22 & -18.37$\pm$0.34\\
J155729.9-225843 & 8.45$\pm$0.93 & 0.28$\pm$0.29 & 8.17$\pm$0.97 & 122.3$\pm$
14.6 & -6.393$\pm$0.38 & -13.91$\pm$0.74 \\
Usco-155744.9-222351 & 9.31$\pm$1.1 & 0.22$\pm$0.25 & 9.09$\pm$1.1 & 110.0$\pm$
13.3 & 6.862$\pm$0.22 & 20.43$\pm$0.36\\
J155829.8-231007 & 6.96$\pm$0.67 & -0.26$\pm$0.11 & 7.22$\pm$0.68 & 138.5$\pm$
13.0 & -8.736$\pm$0.11 & -18.64$\pm$0.14  \\
J155918.4-221042 & 8.08$\pm$1.6 & -0.22$\pm$0.20 & 8.30$\pm$1.6 & 120.5$\pm$23.0
 & -6.565$\pm$0.17 & -19.64$\pm$0.27 \\
USco-160004.3-223014 & 9.85$\pm$7.0 & 1.2$\pm$0.68 & 8.60$\pm$7.0 & 116.3$\pm$
94.7 & -133.3$\pm$3.1 & 40.45$\pm$2.7 \\
J160013.3-241810 & 5.39$\pm$1.5 & -0.26$\pm$0.14 & 5.65$\pm$1.5 & 177.0$\pm$47.7
 & -7.530$\pm$0.58 & -20.56$\pm$0.68 \\
Usco-160018.4-223011 & 10.1$\pm$1.2 & 0.11$\pm$0.28 & 9.99$\pm$1.3 & 100.1$\pm$
12.5 & -5.697$\pm$0.23 & -17.77$\pm$0.29 \\
J160108.0-211318 & 6.60$\pm$1.0 & -0.39$\pm$0.15 & 6.99$\pm$1.0 & 143.0$\pm$21.3
 & -9.407$\pm$0.13 & -18.62$\pm$0.19 \\
J160159.7-195219 & 10.1$\pm$0.42 & -0.20$\pm$0.19 & 10.2$\pm$0.46 & 97.58$\pm$
4.38 & -18.16$\pm$0.15 & -26.21$\pm$0.27 \\
J160200.3-222123 & 5.55$\pm$0.41 & 0.0083$\pm$0.25 & 5.54$\pm$0.48 & 180.5$\pm$
15.7 & -8.808$\pm$0.19 & -20.46$\pm$0.79 \\
USco-160202.9-223613 & 3.96$\pm$1.1 & 0.13$\pm$0.27 & 3.83$\pm$1.1 & 260.8$\pm$
77.2 & -11.46$\pm$0.47 & -29.50$\pm$0.43 \\
J160226.2-200241 & 5.98$\pm$1.0 & -0.51$\pm$0.18 & 6.49$\pm$1.1 & 154.1$\pm$25.1
 & -5.486$\pm$0.14 & -20.77$\pm$0.26 \\
Usco-160258.5-225649 & 6.07$\pm$0.33 & 0.059$\pm$0.18 & 6.01$\pm$0.38 & 166.4
$\pm$10.4 & -4.554$\pm$0.12 & -20.04$\pm$0.25 \\
Usco-160325.6-194438 & 13.0$\pm$2.4 & -0.10$\pm$0.14 & 13.1$\pm$2.4 & 76.57$\pm$
14.2 & -60.54$\pm$0.19 & 24.74$\pm$0.25 \\
J160329.4-195503 & 6.41$\pm$0.67 & -0.14$\pm$0.17 & 6.55$\pm$0.69 & 152.7$\pm$
16.1 & -4.972$\pm$0.23 & -16.99$\pm$0.34\\
J160341.8-200557 & 6.45$\pm$0.57 & 0.20$\pm$0.25 & 6.25$\pm$0.62 & 160.0$\pm$
15.9 & -7.249$\pm$0.20 & -18.24$\pm$0.25  \\
J160357.6-203105 & 7.88$\pm$0.55 & -0.38$\pm$0.18 & 8.26$\pm$0.58 & 121.1$\pm$
8.48 & -8.869$\pm$0.39 & -16.79$\pm$0.64  \\
J160357.9-194210 & 6.22$\pm$0.47 & -0.28$\pm$0.21 & 6.50$\pm$0.52 & 154.0$\pm$
12.2 & -6.905$\pm$0.15 & -18.17$\pm$0.26 \\
J160418.2-191055 & 5.95$\pm$0.74 & -0.13$\pm$0.15 & 6.08$\pm$0.76 & 164.5$\pm$
20.4 & -6.289$\pm$0.15 & -17.75$\pm$0.31 \\
J160421.7-213028 & 6.25$\pm$0.65 & -0.17$\pm$0.20 & 6.42$\pm$0.68 & 155.7$\pm$
16.5 & -8.562$\pm$0.18 & -19.97$\pm$0.21  \\
J160435.6-194830 & 8.36$\pm$0.99 & -0.30$\pm$0.15 & 8.66$\pm$1.0 & 115.4$\pm$
13.3 & -5.459$\pm$0.18 & -17.75$\pm$0.21 \\
J160439.1-194245 & 6.32$\pm$0.47 & -0.63$\pm$0.15 & 6.95$\pm$0.49 & 143.9$\pm$
10.2 & -5.141$\pm$0.33 & -18.43$\pm$0.68  \\
J160449.9-203835 & 7.72$\pm$1.3 & -0.062$\pm$0.25 & 7.78$\pm$1.3 & 128.5$\pm$
22.0 & -7.716$\pm$0.16 & -18.56$\pm$0.32\\
J160516.1-193830 & 7.53$\pm$1.7 & -0.12$\pm$0.18 & 7.65$\pm$1.7 & 130.8$\pm$29.9
 & -5.309$\pm$0.27 & -13.22$\pm$0.32 \\
J160532.1-193315 & 7.07$\pm$1.7 & -0.55$\pm$0.12 & 7.61$\pm$1.7 & 131.3$\pm$29.4
 & -6.066$\pm$0.26 & -13.66$\pm$0.29  \\
J160545.4-202308 & 8.95$\pm$1.1 & -0.17$\pm$0.18 & 9.12$\pm$1.2 & 109.6$\pm$14.0
 & -9.630$\pm$0.31 & -20.20$\pm$0.39 \\
J160612.5-203647 & 5.89$\pm$0.78 & -0.087$\pm$0.20 & 5.98$\pm$0.80 & 167.3$\pm$
22.5 & -6.481$\pm$0.21 & -20.56$\pm$0.23  \\
J160622.8-201124 & 7.57$\pm$0.58 & 0.11$\pm$0.22 & 7.46$\pm$0.62 & 134.1$\pm$
11.1 & -5.308$\pm$0.17 & -17.95$\pm$0.39\\
J160643.8-190805 & 4.20$\pm$1.1 & -0.16$\pm$0.22 & 4.36$\pm$1.1 & 229.1$\pm$57.3
 & -5.165$\pm$0.19 & -16.71$\pm$0.31 \\
J160703.9-191132 & 4.98$\pm$2.3 & -0.28$\pm$0.29 & 5.26$\pm$2.3 & 190.1$\pm$81.9
 & -5.965$\pm$0.12 & -20.96$\pm$0.19 \\
J160708.7-192733 & 7.64$\pm$0.44 & 0.18$\pm$0.21 & 7.46$\pm$0.48 & 134.0$\pm$
8.68 & -9.281$\pm$0.47 & -17.95$\pm$1.1\\
J160739.4-191747 & 4.54$\pm$2.1 & 0.89$\pm$0.53 & 3.65$\pm$2.1 & 274.0$\pm$161.
 & -7.392$\pm$2.0 & -22.71$\pm$3.6\\
J160823.2-193001 & 8.76$\pm$0.71 & -0.30$\pm$0.17 & 9.06$\pm$0.73 & 110.3$\pm$
8.87 & -8.837$\pm$0.28 & -18.88$\pm$0.44 \\
J160827.5-194904 & 9.25$\pm$1.7 & 0.21$\pm$0.26 & 9.04$\pm$1.7 & 110.7$\pm$21.2
 & -5.432$\pm$0.43 & -21.99$\pm$0.59 \\
J16083646-24453053 & 9.20$\pm$0.54 & -0.12$\pm$0.26 & 9.32$\pm$0.60 & 107.3$\pm$
6.86 & -11.24$\pm$0.57 & -23.45$\pm$2.7 \\
J160856.7-203346 & 7.16$\pm$0.97 & -0.31$\pm$0.18 & 7.47$\pm$0.99 & 133.9$\pm$
17.7 & -5.086$\pm$0.27 & -21.30$\pm$0.30 \\
J160900.0-190836 & 7.25$\pm$0.78 & -0.21$\pm$0.19 & 7.46$\pm$0.80 & 134.1$\pm$
14.4 & -4.493$\pm$0.58 & -14.79$\pm$1.1 \\
J160900.7-190852 & 7.74$\pm$1.1 & 0.50$\pm$0.34 & 7.24$\pm$1.1 & 138.1$\pm$21.2
 & -3.942$\pm$0.57 & -18.20$\pm$0.92 \\
J160953.6-175446 & 5.06$\pm$1.6 & 0.092$\pm$0.31 & 4.97$\pm$1.7 & 201.3$\pm$67.3
 & -6.265$\pm$0.48 & -18.78$\pm$0.58 \\
J160954.4-190654 & 8.16$\pm$1.0 & 0.50$\pm$0.45 & 7.65$\pm$1.1 & 130.6$\pm$18.7
 & -9.076$\pm$0.81 & -18.98$\pm$0.56 \\
J160959.4-180009 & 7.91$\pm$1.3 & -0.40$\pm$0.12 & 8.31$\pm$1.3 & 120.4$\pm$19.1
 & -4.387$\pm$0.19 & -22.38$\pm$0.29 \\
J161011.0-194603 & 11.0$\pm$0.63 & -0.12$\pm$0.21 & 11.1$\pm$0.66 & 89.93$\pm$
5.38 & -7.300$\pm$0.46 & -18.15$\pm$0.56 \\
J161014.7-191909 & 13.0$\pm$1.3 & 2.0$\pm$0.94 & 11.0$\pm$1.6 & 90.93$\pm$13.0
 & -6.030$\pm$0.50 & -23.12$\pm$1.0 \\
J161024.7-191407 & 5.89$\pm$0.71 & 0.79$\pm$0.73 & 5.10$\pm$1.0 & 196.3$\pm$39.2
 & -8.294$\pm$0.78 & -15.18$\pm$0.77\\
J161052.4-193734 & 7.83$\pm$0.99 & -0.40$\pm$0.16 & 8.23$\pm$1.0 & 121.5$\pm$
14.8 & -5.573$\pm$0.20 & -20.89$\pm$0.20 \\
J161115.3-175721 & 7.45$\pm$1.4 & -0.47$\pm$0.099 & 7.92$\pm$1.4 & 126.2$\pm$
22.2 & -5.482$\pm$0.12 & -20.48$\pm$0.15 \\
J16211564-24361173 & 7.97$\pm$1.1 & 0.52$\pm$0.64 & 7.45$\pm$1.3 & 134.3$\pm$
23.3 & -26.84$\pm$0.40 & -16.73$\pm$0.51 \\
J16212490-24261446 & 4.84$\pm$2.5 & -0.070$\pm$0.88 & 4.91$\pm$2.7 & 203.7$\pm$
111. & -6.432$\pm$0.53 & -16.20$\pm$0.56 \\

\enddata
\label{tab:para}
\tablecomments{See Section~\ref{sub:data} for more details}
\end{deluxetable}

\begin{deluxetable}{llrrrrrrrr}
\tablewidth{0pt}
\tabletypesize{\footnotesize}
\tablecaption{Parameters Used for Isochronal Ages}
\tablehead{
\colhead{Designation} & \colhead{Spectral} & \colhead{Ref} & \colhead{T$_{eff}$} & \colhead{A$_V$} &
	\colhead{Ref} & \colhead{H$_{abs}$} & \colhead{Age} & \colhead{Excess} & \colhead{Ref} \\
	&\colhead{Type} & & \colhead{(K)} & && \colhead{(mag)} & \colhead{(Myr)} \\}
\startdata
J155106.6-240218 & M2 & 1 &       3490.00$\pm$
      135.000 &0.40 & 1 & 4.14$\pm$0.17 & $6\pm4$ &        N & 1\\
J155624.8-222555 & M4 & 2 &       3160.00$\pm$      72.0000 &1.7 & 
1 & 3.95$\pm$0.13 &$6\pm3$  &        Y & 1\\
J155655.5-225839 & M0.5 & 3 &       3700.00$\pm$      70.0000 &0.10 & 
3 & 3.78$\pm$0.43 & $5\pm5$ &        N & 3\\
J155706.4-220606 & M4 & 2 &       3160.00$\pm$      72.0000 &2.0 & 
1 & 4.07$\pm$0.25 & $7\pm5$ &        Y & 1\\
J155729.9-225843 & M4 & 2 &       3160.00$\pm$      72.0000 &1.4 & 
1 & 4.66$\pm$0.26 & $7\pm5$ &        Y & 1\\
Usco-155744.9-222351 & M2 & 2 &       3490.00$\pm$      40.5000 &0.70
 & 2 & 5.11$\pm$0.26 & $29\pm20$ &        ? & --\\
J155829.8-231007 & M3 & 2 &       3360.00$\pm$      49.5000 &1.3 & 
1 & 4.66$\pm$0.20 & $16\pm10$ &        Y & 1\\
J155918.4-221042 & M4 & 2 &       3160.00$\pm$      72.0000 &1.3 & 
1 & 3.64$\pm$0.41 & $2\pm2$ &        N & 1\\
USco-160004.3-223014 & M3 & 2 &       3360.00$\pm$      49.5000 &0.20
 & 2 & 6.78$\pm$1.8 & $89\pm44$ &        ? & --\\
J160013.3-241810 & M0.5 & 3 &       3700.00$\pm$      70.0000 &0.70 & 
9 & 2.79$\pm$0.59 & $3\pm4$ &        Y & 9\\
Usco-160018.4-223011 & M4.5 & 4 &       3020.00$\pm$      210.000 &0.70
 & 9 & 5.18$\pm$0.27 & $4\pm2$ &        Y & 9\\
J160108.0-211318 & K7 & 5 &       3970.00$\pm$      20.0000 &0.0 & 
1 & 3.25$\pm$0.32 & $4\pm3$ &        N & 1\\
J160159.7-195219 & M5 & 6 &       2880.00$\pm$      90.0000 &0.60 & 
1 & 6.27$\pm$0.097 & $5\pm3$ &        N & 1\\
J160200.3-222123 & M1.0 & 3 &       3630.00$\pm$      70.0000 &0.70 & 
9 & 2.12$\pm$0.19 & $15\pm10$ &        Y & 9\\
USco-160202.9-223613 & M0 & 2 &       3770.00$\pm$      42.0000 &0.75
 & 9 & 4.13$\pm$0.64 & $20\pm23$ &        Y & 9\\
J160226.2-200241 & M5 & 6 &       2880.00$\pm$      90.0000 &0.30 & 
1 & 5.01$\pm$0.35 & $1\pm1$ &        N & 1\\
Usco-160258.5-225649 & M2.75 & 4 &       3392.50$\pm$      157.500 &0.80
 & 2 & 3.45$\pm$0.14 & $3\pm2$ &        N & 4\\
Usco-160325.6-194438 & M2 & 2 &       3490.00$\pm$      40.5000 &1.6
 & 2 & 5.68$\pm$0.40 & $79\pm43$ &        ? & --\\
J160329.4-195503 & M5 & 6 &       2880.00$\pm$      90.0000 &0.30 & 
1 & 5.08$\pm$0.23 & $1\pm1$ &        N & 1\\
J160341.8-200557 & M2 & 6 &       3490.00$\pm$      67.5000 &0.90 & 
1 & 2.84$\pm$0.22 & $1\pm1$ &        N & 1\\
J160357.6-203105 & K5 & 4 &       4140.00$\pm$      155.000 &0.70 & 
9 & 2.66$\pm$0.15 & $7\pm3$ &        Y & 9\\
J160357.9-194210 & M2 & 6 &       3490.00$\pm$      67.5000 &0.70 & 
9 & 3.97$\pm$0.17 & $7\pm4$ &        Y & 9\\
J160418.2-191055 & M4 & 6 &       3160.00$\pm$      120.000 &0.80 & 
1 & 4.13$\pm$0.27 &$3\pm2$  &        N & 1\\
J160421.7-213028 & K2 & 1 &       4760.00$\pm$
      185.000 &0.70 & 9 & 2.44$\pm$0.23 & $19\pm13$ &        Y & 9\\
J160435.6-194830 & M5.25 & 4 &       2860.00$\pm$      163.750 &0.80 & 
1 & 5.37$\pm$0.25 & $3\pm2$ &        N & 1\\
J160439.1-194245 & M3.25 & 4 &       3310.00$\pm$      183.750 &0.37 & 
9 & 4.85$\pm$0.15 & $8\pm4$ &        Y & 9\\
J160449.9-203835 & M5 & 6 &       2880.00$\pm$      90.0000 &0.70 & 
1 & 5.03$\pm$0.37 &$2\pm1$  &        N & 1\\
J160516.1-193830 & M4.5 & 4 &       3020.00$\pm$      210.000 &0.70 & 
1 & 5.43$\pm$0.50 &$5\pm5$  &        N & 1\\
J160532.1-193315 & M4.75 & 4 &       2950.00$\pm$      195.000 &0.20 & 
9 & 6.10$\pm$0.49 & $5\pm6$ &        Y & 9\\
J160545.4-202308 & M2 & 6 &       3490.00$\pm$      67.5000 &1.6 & 
9 & 3.94$\pm$0.28 & $15\pm11$ &        Y & 9\\
J160612.5-203647 & K5 & 1 &       4140.00$\pm$
      155.000 &1.8 & 1 & 1.22$\pm$0.29 & $3\pm3$ &        N & 1\\
J160622.8-201124 & M5 & 6 &       2880.00$\pm$      90.0000 &0.20 & 
1 & 5.44$\pm$0.18 & $1\pm1$ &        Y & 1\\
J160643.8-190805 & K6 & 6 &       4020.00$\pm$      42.5000 &0.75 & 
9 & 1.85$\pm$0.54 &$1\pm6$  &        Y & 9\\
J160703.9-191132 & M1 & 1 &       3630.00$\pm$
      140.000 &1.1 & 1 & 2.03$\pm$0.94 & $1\pm31$ &        N & 1\\
J160708.7-192733 & M4 & 6 &       3160.00$\pm$      120.000 &1.1 & 
9 & 4.67$\pm$0.14 &$6\pm4$  &        Y & 9\\
J160739.4-191747 & M2 & 6 &       3490.00$\pm$      67.5000 &0.76 & 
9 & 2.14$\pm$1.3 & $10\pm32$ &        Y & 9\\
J160823.2-193001 & K9 & 6 &       3880.00$\pm$      30.0000 &0.70 & 
9 & 3.82$\pm$0.17 &$19\pm11$  &        Y & 9\\
J160827.5-194904 & M5 & 6 &       2880.00$\pm$      90.0000 &0.72 & 
9 & 5.04$\pm$0.42 & $2\pm1$ &        Y & 9\\
J16083646-24453053 & M3.5 & 4 &       3260.00$\pm$      202.500 &-- & 
-- & 5.79$\pm$0.14 & $11\pm6$ &        N & 4\\
J160856.7-203346 & K5 & 7 &       4140.00$\pm$      155.000 &1.4 & 
1 & 1.74$\pm$0.29 & $5\pm3$ &        N & 1\\
J160900.0-190836 & M5 & 2 &       2880.00$\pm$      54.0000 &0.31 & 
9 & 5.39$\pm$0.23 & $2\pm15$ &        Y & 9\\
J160900.7-190852 & K9 & 6 &       3880.00$\pm$      30.0000 &0.70 & 
9 & 3.07$\pm$0.33 & $6\pm5$ &        Y & 9\\
J160953.6-175446 & M3 & 2 &       3360.00$\pm$      49.5000 &1.7 & 
9 & 3.63$\pm$0.73 & $7\pm7$ &        Y & 9\\
J160954.4-190654 & M1 & 6 &       3630.00$\pm$      70.0000 &0.70 & 
9 & 3.59$\pm$0.31 & $6\pm5$ &        Y & 9\\
J160959.4-180009 & M4 & 6 &       3160.00$\pm$      120.000 &0.58 & 
9 & 4.64$\pm$0.34 & $4\pm3$ &        Y & 9\\
J161011.0-194603 & M5 & 2 &       2880.00$\pm$      54.0000 &0.50 & 
1 & 6.44$\pm$0.13 & $6\pm6$ &        Y & 1\\
J161014.7-191909 & M2 & 6 &       3490.00$\pm$      67.5000 &0.87 & 
9 & 4.67$\pm$0.31 & $20\pm15$ &        Y & 9\\
J161024.7-191407 & M3 & 6 &       3360.00$\pm$      82.5000 &1.5 & 
1 & 2.70$\pm$0.43 & $1\pm10$ &        N & 1\\
J161052.4-193734 & M3 & 2 &       3360.00$\pm$      49.5000 &2.3 & 
1 & 3.34$\pm$0.26 & $9\pm6$ &        N & 1\\
J161115.3-175721 & M1 & 6 &       3630.00$\pm$      70.0000 &0.70 & 
9 & 3.28$\pm$0.38 & $4\pm3$ &        Y & 9\\
J16211564-24361173 & M3.5 & 8 &       3260.00$\pm$      100.000 &-- & 
-- & 6.18$\pm$0.38 & $17\pm12$ &        N & 4\\
J16212490-24261446 & M3.5 & 8 &       3260.00$\pm$      100.000 &-- & 
-- & 5.32$\pm$1.2 & $15\pm15$ &        N & 4\\
\enddata
\label{tab:ages}
\tablerefs{(1)~\cite{Carpenter09}, (2)~\cite{Preibisch02}, (3)~\cite{Rizzuto15}, (4)~\cite{Luhman12}, (5)~\cite{Riaz06}, 
(6)~\cite{Preibisch01}, (7)~\cite{Preibisch98}, (8)~\cite{Slesnick08}, (9)~\cite{Barenfeld16}
}
\tablecomments{See Section~\ref{sub:ages} for more details}
\end{deluxetable}


\begin{thebibliography}{}

\bibitem[\protect\citeauthoryear{{Aarnio} et~al.}{{Aarnio}
  et~al.}{2008}]{Aarnio08}
{Aarnio}, A.~N., {Weinberger}, A.~J., {Stassun}, K.~G., {Mamajek}, E.~E.,  \&
  {James}, D.~J. 2008, \aj, 136, 2483

\bibitem[\protect\citeauthoryear{{Adams}}{{Adams}}{2000}]{Adams00}
{Adams}, F.~C. 2000, \apj, 542, 964

\bibitem[\protect\citeauthoryear{{Anglada-Escud{\'e}}
  et~al.}{{Anglada-Escud{\'e}} et~al.}{2012}]{Anglada12}
{Anglada-Escud{\'e}}, G., {Boss}, A.~P., {Weinberger}, A.~J., {Thompson},
  I.~B., {Butler}, R.~P., {Vogt}, S.~S.,  \& {Rivera}, E.~J. 2012, \apj, 746,
  37

\bibitem[\protect\citeauthoryear{{Balog} et~al.}{{Balog}
  et~al.}{2007}]{Balog07}
{Balog}, Z., {Muzerolle}, J., {Rieke}, G.~H., {Su}, K.~Y.~L., {Young}, E.~T.,
  \& {Megeath}, S.~T. 2007, \apj, 660, 1532

\bibitem[\protect\citeauthoryear{{Baraffe}, {Chabrier}, \&
  {Gallardo}}{{Baraffe} et~al.}{2009}]{Baraffe09}
{Baraffe}, I., {Chabrier}, G.,  \& {Gallardo}, J. 2009, \apjl, 702, L27

\bibitem[\protect\citeauthoryear{{Baraffe} et~al.}{{Baraffe}
  et~al.}{2015}]{Baraffe15}
{Baraffe}, I., {Homeier}, D., {Allard}, F.,  \& {Chabrier}, G. 2015, \aap, 577,
  A42

\bibitem[\protect\citeauthoryear{{Barenfeld} et~al.}{{Barenfeld}
  et~al.}{2016}]{Barenfeld16}
{Barenfeld}, S.~A., {Carpenter}, J.~M., {Ricci}, L.,  \& {Isella}, A. 2016,
  \apj, 827, 142

\bibitem[\protect\citeauthoryear{{Boss} et~al.}{{Boss} et~al.}{2009}]{Boss09}
{Boss}, A.~P., et~al. 2009, \pasp, 121, 1218

\bibitem[\protect\citeauthoryear{{Bouy} et~al.}{{Bouy} et~al.}{2006}]{Buoy06}
{Bouy}, H., {Mart{\'{\i}}n}, E.~L., {Brandner}, W., {Zapatero-Osorio}, M.~R.,
  {B{\'e}jar}, V.~J.~S., {Schirmer}, M., {Hu{\'e}lamo}, N.,  \& {Ghez}, A.~M.
  2006, \aap, 451, 177

\bibitem[\protect\citeauthoryear{{Cambr{\'e}sy} et~al.}{{Cambr{\'e}sy}
  et~al.}{2002}]{Cambresy02}
{Cambr{\'e}sy}, L., {Beichman}, C.~A., {Jarrett}, T.~H.,  \& {Cutri}, R.~M.
  2002, \aj, 123, 2559

\bibitem[\protect\citeauthoryear{{Carpenter} et~al.}{{Carpenter}
  et~al.}{2006}]{Carpenter06}
{Carpenter}, J.~M., {Mamajek}, E.~E., {Hillenbrand}, L.~A.,  \& {Meyer}, M.~R.
  2006, \apjl, 651, L49

\bibitem[\protect\citeauthoryear{{Carpenter} et~al.}{{Carpenter}
  et~al.}{2009}]{Carpenter09}
{Carpenter}, J.~M., {Mamajek}, E.~E., {Hillenbrand}, L.~A.,  \& {Meyer}, M.~R.
  2009, \apj, 705, 1646

\bibitem[\protect\citeauthoryear{{Carpenter}, {Ricci}, \& {Isella}}{{Carpenter}
  et~al.}{2014}]{Carpenter14}
{Carpenter}, J.~M., {Ricci}, L.,  \& {Isella}, A. 2014, \apj, 787, 42

\bibitem[\protect\citeauthoryear{{Cohen} \& {Kuhi}}{{Cohen} \&
  {Kuhi}}{1979}]{Cohen79}
{Cohen}, M.,  \& {Kuhi}, L.~V. 1979, \apjl, 227, L105

\bibitem[\protect\citeauthoryear{{Dahm}, {Slesnick}, \& {White}}{{Dahm}
  et~al.}{2012}]{Dahm12}
{Dahm}, S.~E., {Slesnick}, C.~L.,  \& {White}, R.~J. 2012, \apj, 745, 56

\bibitem[\protect\citeauthoryear{{de Zeeuw} et~al.}{{de Zeeuw}
  et~al.}{1999}]{deZeeuw99}
{de Zeeuw}, P.~T., {Hoogerwerf}, R., {de Bruijne}, J.~H.~J., {Brown}, A.~G.~A.,
   \& {Blaauw}, A. 1999, \aj, 117, 354

\bibitem[\protect\citeauthoryear{{Donaldson} et~al.}{{Donaldson}
  et~al.}{2016}]{Donaldson16}
{Donaldson}, J.~K., {Weinberger}, A.~J., {Gagn{\'e}}, J., {Faherty}, J.~K.,
  {Boss}, A.~P.,  \& {Keiser}, S.~A. 2016, \apj, 833, 95

\bibitem[\protect\citeauthoryear{{Feiden}}{{Feiden}}{2016}]{Feiden16}
{Feiden}, G.~A. 2016, \aap, 593, A99

\bibitem[\protect\citeauthoryear{{Gagn{\'e}} et~al.}{{Gagn{\'e}}
  et~al.}{2017}]{Gagne17}
{Gagn{\'e}}, J., et~al. 2017, \apjs, 228, 18

\bibitem[\protect\citeauthoryear{{Gagn{\'e}} et~al.}{{Gagn{\'e}}
  et~al.}{2014}]{Gagne14}
{Gagn{\'e}}, J., {Lafreni{\`e}re}, D., {Doyon}, R., {Malo}, L.,  \& {Artigau},
  {\'E}. 2014, \apj, 783, 121

\bibitem[\protect\citeauthoryear{{Kraus} \& {Hillenbrand}}{{Kraus} \&
  {Hillenbrand}}{2007}]{Kraus07}
{Kraus}, A.~L.,  \& {Hillenbrand}, L.~A. 2007, \apj, 662, 413

\bibitem[\protect\citeauthoryear{{Luhman} \& {Mamajek}}{{Luhman} \&
  {Mamajek}}{2012}]{Luhman12}
{Luhman}, K.~L.,  \& {Mamajek}, E.~E. 2012, \apj, 758, 31

\bibitem[\protect\citeauthoryear{{Malo} et~al.}{{Malo} et~al.}{2013}]{Malo13}
{Malo}, L., {Doyon}, R., {Lafreni{\`e}re}, D., {Artigau}, {\'E}., {Gagn{\'e}},
  J., {Baron}, F.,  \& {Riedel}, A. 2013, \apj, 762, 88

\bibitem[\protect\citeauthoryear{{Meyer} et~al.}{{Meyer}
  et~al.}{2007}]{Meyer07}
{Meyer}, M.~R., {Backman}, D.~E., {Weinberger}, A.~J.,  \& {Wyatt}, M.~C. 2007,
  Protostars and Planets V, 573

\bibitem[\protect\citeauthoryear{{Olczak}, {Pfalzner}, \& {Eckart}}{{Olczak}
  et~al.}{2010}]{Olczak10}
{Olczak}, C., {Pfalzner}, S.,  \& {Eckart}, A. 2010, \aap, 509, A63

\bibitem[\protect\citeauthoryear{{Pecaut} \& {Mamajek}}{{Pecaut} \&
  {Mamajek}}{2013}]{Pecaut13}
{Pecaut}, M.~J.,  \& {Mamajek}, E.~E. 2013, \apjs, 208, 9

\bibitem[\protect\citeauthoryear{{Pecaut} \& {Mamajek}}{{Pecaut} \&
  {Mamajek}}{2016}]{Pecaut16}
{Pecaut}, M.~J.,  \& {Mamajek}, E.~E. 2016, \mnras, 461, 794

\bibitem[\protect\citeauthoryear{{Pecaut}, {Mamajek}, \& {Bubar}}{{Pecaut}
  et~al.}{2012}]{Pecaut12}
{Pecaut}, M.~J., {Mamajek}, E.~E.,  \& {Bubar}, E.~J. 2012, \apj, 746, 154

\bibitem[\protect\citeauthoryear{{Preibisch} et~al.}{{Preibisch}
  et~al.}{2002}]{Preibisch02}
{Preibisch}, T., {Brown}, A.~G.~A., {Bridges}, T., {Guenther}, E.,  \&
  {Zinnecker}, H. 2002, \aj, 124, 404

\bibitem[\protect\citeauthoryear{{Preibisch}, {Guenther}, \&
  {Zinnecker}}{{Preibisch} et~al.}{2001}]{Preibisch01}
{Preibisch}, T., {Guenther}, E.,  \& {Zinnecker}, H. 2001, \aj, 121, 1040

\bibitem[\protect\citeauthoryear{{Preibisch} et~al.}{{Preibisch}
  et~al.}{1998}]{Preibisch98}
{Preibisch}, T., {Guenther}, E., {Zinnecker}, H., {Sterzik}, M., {Frink}, S.,
  \& {Roeser}, S. 1998, \aap, 333, 619

\bibitem[\protect\citeauthoryear{{Preibisch} \& {Mamajek}}{{Preibisch} \&
  {Mamajek}}{2008}]{Preibisch08}
{Preibisch}, T.,  \& {Mamajek}, E. 2008, {The Nearest OB Association:
  Scorpius-Centaurus (Sco OB2)}, ed. B.~{Reipurth} 235

\bibitem[\protect\citeauthoryear{{Preibisch} \& {Zinnecker}}{{Preibisch} \&
  {Zinnecker}}{1999}]{Preibisch99}
{Preibisch}, T.,  \& {Zinnecker}, H. 1999, \aj, 117, 2381

\bibitem[\protect\citeauthoryear{{Riaz}, {Gizis}, \& {Harvin}}{{Riaz}
  et~al.}{2006}]{Riaz06}
{Riaz}, B., {Gizis}, J.~E.,  \& {Harvin}, J. 2006, \aj, 132, 866

\bibitem[\protect\citeauthoryear{{Riedel} et~al.}{{Riedel}
  et~al.}{2017}]{Riedel17}
{Riedel}, A.~R., {Blunt}, S.~C., {Lambrides}, E.~L., {Rice}, E.~L., {Cruz},
  K.~L.,  \& {Faherty}, J.~K. 2017, \aj, 153, 95

\bibitem[\protect\citeauthoryear{{Rizzuto}, {Ireland}, \& {Kraus}}{{Rizzuto}
  et~al.}{2015}]{Rizzuto15}
{Rizzuto}, A.~C., {Ireland}, M.~J.,  \& {Kraus}, A.~L. 2015, \mnras, 448, 2737

\bibitem[\protect\citeauthoryear{{Robin} et~al.}{{Robin}
  et~al.}{2012}]{Robin2012}
{Robin}, A.~C., {Marshall}, D.~J., {Schultheis}, M.,  \& {Reyl{\'e}}, C. 2012,
  \aap, 538, A106

\bibitem[\protect\citeauthoryear{{Slesnick}, {Hillenbrand}, \&
  {Carpenter}}{{Slesnick} et~al.}{2008}]{Slesnick08}
{Slesnick}, C.~L., {Hillenbrand}, L.~A.,  \& {Carpenter}, J.~M. 2008, \apj,
  688, 377

\bibitem[\protect\citeauthoryear{{Song}, {Zuckerman}, \& {Bessell}}{{Song}
  et~al.}{2012}]{Song12}
{Song}, I., {Zuckerman}, B.,  \& {Bessell}, M.~S. 2012, \aj, 144, 8

\bibitem[\protect\citeauthoryear{{Walter} et~al.}{{Walter}
  et~al.}{1994}]{Walter94}
{Walter}, F.~M., {Vrba}, F.~J., {Mathieu}, R.~D., {Brown}, A.,  \& {Myers},
  P.~C. 1994, \aj, 107, 692

\bibitem[\protect\citeauthoryear{{Weinberger}, {Anglada-Escud{\'e}}, \&
  {Boss}}{{Weinberger} et~al.}{2013}]{Weinberger13}
{Weinberger}, A.~J., {Anglada-Escud{\'e}}, G.,  \& {Boss}, A.~P. 2013, \apj,
  762, 118

\bibitem[\protect\citeauthoryear{{Weinberger} et~al.}{{Weinberger}
  et~al.}{2016}]{Weinberger16}
{Weinberger}, A.~J., {Boss}, A.~P., {Keiser}, S.~A., {Anglada-Escud{\'e}}, G.,
  {Thompson}, I.~B.,  \& {Burley}, G. 2016, \aj, 152, 24

\bibitem[\protect\citeauthoryear{{Zacharias} et~al.}{{Zacharias}
  et~al.}{2013}]{Zacharias13}
{Zacharias}, N., {Finch}, C.~T., {Girard}, T.~M., {Henden}, A., {Bartlett},
  J.~L., {Monet}, D.~G.,  \& {Zacharias}, M.~I. 2013, \aj, 145, 44

\end{thebibliography}
\end{document}